\documentclass{sig-alternate-10pt}

\pdfoutput=1
\usepackage[margin=1in]{geometry}
\usepackage[utf8]{inputenc}
\usepackage{subfigure}
\usepackage[dvipsnames]{xcolor}
\usepackage[hyphens]{url}
\usepackage[hyperref]{ntheorem}
\usepackage{mdframed}
\usepackage[inline]{enumitem}
\usepackage{wrapfig}
\usepackage{tikz}
\usepackage{listings}
\usepackage{caption}
\usepackage{verbatim}

\usetikzlibrary{positioning}

\definecolor{commentsColor}{rgb}{0.497495, 0.497587, 0.497464}
\definecolor{keywordsColor}{rgb}{0.000000, 0.000000, 0.635294}
\definecolor{stringColor}{rgb}{0.558215, 0.000000, 0.135316}
\definecolor{backColour}{rgb}{0.95,0.95,0.92}
\newcounter{enum}
\newenvironment{packed_enum}{
\begin{list}{\textbf{(\arabic{enum})}}{
 \setlength{\topsep}{0pt}
 \setlength{\itemsep}{0.5mm}
 \setlength{\parskip}{0pt}
 \setlength{\parsep}{0pt}
 \setlength{\labelwidth}{-5 pt}
 \setlength{\leftmargin}{0 pt}
 \setlength{\itemindent}{0pt}
 \usecounter{enum}}
}{\end{list}}

\newenvironment{packed_item}{
\begin{list}{\textbf{$\bullet$}}{
 \setlength{\itemsep}{0.5mm}
 \setlength{\parskip}{-2mm}
 \setlength{\parsep}{0pt}
 \setlength{\labelwidth}{-5 pt}
 \setlength{\leftmargin}{0 pt}
 \setlength{\itemindent}{0pt}}
}{\end{list}}

\theoremstyle{break}
\theoremheaderfont{\bfseries}
\newmdtheoremenv[
 linecolor=white,
 leftmargin=0,
 rightmargin=0,
 backgroundcolor=gray!10,
 innertopmargin=3pt,
 ntheorem
]{example}{Example}[section]

\definecolor{myviolet}{RGB}{203,6,166}
\definecolor{mygreen}{RGB}{72,186,59}
\newcommand*\bcircled[1]{\tikz[baseline=(char.base)]{
 \node[shape=circle,fill=black,text=white,inner sep=0.5pt] (char) {\textbf{#1}};}}
\newcommand*\vcircled[1]{\tikz[baseline=(char.base)]{
 \node[shape=circle,fill=myviolet,text=white,inner sep=0.5pt] (char) {\textbf{#1}};}}
\newcommand*\gcircled[1]{\tikz[baseline=(char.base)]{
 \node[shape=circle,fill=mygreen,text=white,inner sep=0.5pt] (char) {\textbf{#1}};}}
\newcommand{\myparagraph}[1]{\vspace{0.1cm}\noindent\textbf{{#1}.~}}
\newcounter{mycounter}
\newcommand{\challenge}[1]{\myparagraph{\stepcounter{mycounter}(RC\themycounter)~#1}}

\begin{document}
  \pagestyle{empty}
  \title{Agora: A Unified Asset Ecosystem Going Beyond Marketplaces and Cloud Services [Vision]}

  \numberofauthors{1}
  \author{
    \alignauthor
    Jonas Traub \hfill Zoi Kaoudi \hfill Kaustubh Beedkar \hfill Sergey Redyuk\\
    \hfill Viktor Rosenfeld \hfill Jorge-Arnulfo Quian\'e-Ruiz \hfill Volker Markl\hfill~\\
    \vspace{4mm}
    \normalsize Technische Universität Berlin \hspace{15mm} DFKI GmbH \hspace{10mm}
  }

  \maketitle
  \sloppypar

  \begin{abstract}
Data, algorithms, and compute/storage infrastructure are key assets that drive data science and artificial intelligence applications.
As providing all these assets requires a huge investment, data science and artificial intelligence technologies are currently dominated by a small number of providers who can afford these investments.
This leads to lock-in effects and hinders features that require a flexible exchange of assets among users.

In this vision paper, we present Agora, a unified asset ecosystem.
The Agora system provides the technical infrastructure that allows for offering and using data and algorithms, as well as physical infrastructure components.
Agora is designed as an open ecosystem of asset marketplaces and provides  to a broad audience not only data but the entire data value chain (including computational resources and human expertise).
Agora (i) leverages a fine-grained exchange of assets,
(ii) allows for combining assets to novel applications, and (iii) flexibly executes such applications on available resources.
As a result, Agora overcomes lock-in effects and removes entry barriers for new asset providers.
In contrast to existing data management systems, Agora operates in a heavily decentralized and dynamic environment:
Data, algorithms, and even compute resources are dynamically created, modified, and removed by different stakeholders.
Agora presents novel research directions for the data management community as a whole: It requires to combine our traditional expertise in scalable data processing and management with infrastructure provisioning as well as economic and application aspects of data, algorithms, and infrastructure.
\vspace{5mm}
\end{abstract}

  \vspace{-2mm}
  \section{Introduction}\label{sec:intro}
  The ongoing digitalization has a profound impact on industry, science, and society as a whole.
The access to data as well as to data science (DS) technology constitute a critical point of control:
Wide access to both of them is crucial for economic success and scientific progress, promoting a new data-centric economy~\cite{theeconomist}.
Nowadays, business leaders talk about the fourth industrial revolution~\cite{klausschwab}.
The fourth paradigm of data-intensive scientific discovery facilitates new insights through the analysis of large datasets that are generated from modern scientific experiments~\cite{4paradigm}.

Data has become a fundamental {\em factor of production}.
In contrast to natural resources like oil, data can be exploited infinitely.
It can be repeatedly curated and analyzed with DS technologies to produce new insights and solve problems in a more efficient way.
Data {\em together} with DS technologies are competitive differentiators in the data economy.
Companies that are proficient at utilizing them grow faster and perform better than their competitors~\cite{oxfordeconomics}.
As a result, the data economy is quickly developing a strong dependency on a small number of DS proficient companies.
This implicitly causes {\em lock-in} effects on customers, which, in turn, might cause customers to use suboptimal solutions or even to not have a solution at all.
\begin{figure*}[t!]
  \centering
  \includegraphics[width=\linewidth]{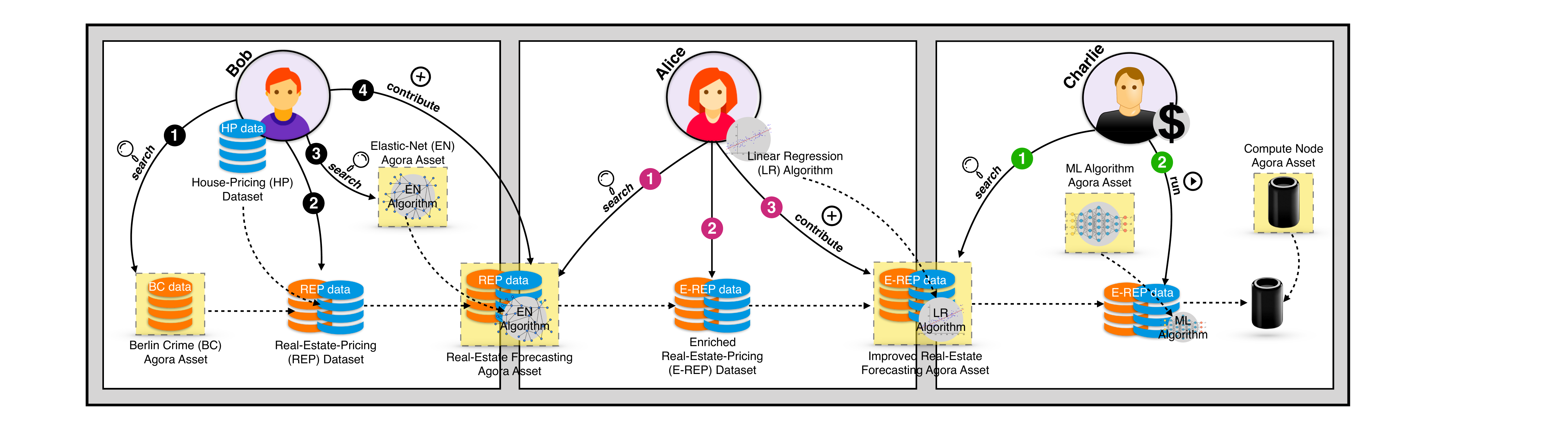}
  \caption{Motivating examples: Bob, Alice, and Charlie use Agora to discover assets, improve them, and contribute them back to the ecosystem. Agora also provides the infrastructure to optimize and run these assets (e.g., in the case of Charlie).}
  \label{fig:usage}
\end{figure*}

\subsection{Towards a Unified Asset Ecosystem}
As data and DS technologies production factors, it is clear that they must be accessible by {\em everyone}.
In fact, the database community has recently recognized that removing such lock-in effects will significantly benefit all users~\cite{seattle-report}.
Academia and industry have made progress towards this goal by providing access to data~\cite{datahub,iota,oceanprotocol}, AI algorithms~\cite{acumos,awsmarketplace,bonseyes,azure}, expertise (services)~\cite{experfy,salesforce}, or computational resources~\cite{enigma}.
However, the users still require significant expertise to combine all these {\em data-related assets} (assets, for short) from different marketplaces and cloud providers.
For instance, a social scientist, who has no expertise in DS techniques and does not own any data, can hardly validate her assumptions about a social phenomenon, even if the required data and technology exists.
We thus need an ecosystem that provides {\em unified} access to all types of assets: (i)~high-quality data, (ii)~state-of-the-art DS technology and expertise, and (iii)~compute and storage resources.
The treatment of these types of assets in a uniform and systematic way allows for easy creation and composition of data science pipelines, both with respect to the algorithmic and data specification as well as its scalable execution.

\subsection{Our Vision: The Agora Ecosystem}
We envision \textit{Agora}, an ecosystem that brings together asset providers and consumers to solve data-related problems using DS.
Agora allows providers to offer any type of assets (e.g., data, algorithms, software, computational resources, human expertise) to a broader audience.
Also, it enables both the experts and the non-expert users to gain insights or enhance their businesses by combining and using the assets.
Ultimately, Agora aims at providing access not only to data sources but to the entire data value chain.

We envision this ecosystem as a two-layer abstraction: the {\em asset layer} and the {\em execution layer}.
The asset layer is composed of a set of marketplaces where providers and consumers can exchange assets.
The execution layer provides the means to users to run their tasks (composition of assets) in Agora instead of using their own computing infrastructure.
The key aspect of Agora is the fine-grained exchange of {\em any} asset.
Each type of assets corresponds to a specialization of the provider, leading to different user roles.
Agora hides the complexity of each role.
For example,
\begin{enumerate*}[label=(\roman*)]
  \item a researcher can subscribe to a stream of events without knowing any detail about the infrastructure that captures those events;
  \item a company can acquire a classification pipeline without understanding the details of all involved algorithms;
  \item researchers and companies can book a stream processing cluster with uptime guarantees without having any knowledge on cluster operations; and
  \item system operators can focus on cluster monitoring and maintenance without knowing any detail about the tasks running on top of the cluster.
\end{enumerate*}

Overall, we see Agora as an umbrella system, which unites all pieces of data management research in an open and collaborative ecosystem.
We, thus, believe that the database community should drive the realization of this vision.

\subsection{Motivating Examples}\label{section:motivation}
Imagine Bob, a freelance data scientist, who wants to create a machine learning (ML) model for real-estate price forecasting in Berlin.
His dataset is missing the criminality rate of each area, which he knows also affects the prices.
He, thus, goes to Agora to find data about the crime rates in Berlin~\bcircled{1}.
He finds the data, augments his initial dataset with this feature~\bcircled{2}, and builds an ML model using the elastic-net algorithm~\bcircled{3}.
He then decides to provide his composed asset in Agora~\bcircled{4}.
Bob's asset consists of the `real-estate-pricing' dataset for Berlin and the elastic-net algorithm to estimate a potential price of apartments.

Alice, another data scientist, finds Bob's asset in Agora~\vcircled{1} and decides to improve it~\vcircled{2}.
She enriches the original `real-estate-pricing' dataset with several feature engineering techniques, adds the `linear-regression' algorithm for prediction, and contributes it back to Agora to gain some revenue~\vcircled{3}.

Charlie, a consumer who is looking for a real-estate pricing predictor, queries Agora for available assets on price forecasting that yield the average error rate below 5,000 euros~\gcircled{1}.
As he does not have the infrastructure to run assets in his home, he decides to use Agora to also execute his discovered assets (e.g.,~train the ML pipelines he has found)~\gcircled{2}.
Although he wants to complete the training as fast as possible, his budget is limited.
To overcome his budget limitation, Agora replaces the linear regression algorithm by a logically equivalent neural network that achieves better performance.
Next, Agora decides to run the resulting asset on an execution node registered as an asset within Agora.

Allowing asset exchange in Agora leads to the following main benefits:

\emph{(1) Secondary use of existing assets.} Users can reuse any (composed) asset (e.g.,~data and algorithms) offered in Agora.
In most cases, companies own a plethora of highly valuable assets.
However, as these assets are fragmented across companies, their economical potential remains unused as secondary asset usage is extremely rare.
A fine-grained asset sharing would allow for combining existing resources to derive new insights and services.

\emph{(2) Leveraging specializations.}
Agora creates an ecosystem of highly specialized providers who provide assets of a very high quality.
Such an ecosystem is comparable with the automotive industry where many companies specialize in certain parts (e.g.,~brakes, tires, or lights), which get combined to one high-quality car.
Specialized providers can only operate efficiently if they can offer their assets without massive overhead.
This enables small and medium-sized companies to offer assets that they would not be able to bring in the market otherwise.
Agora, thus, allows consumers to build complex applications by combining high-quality assets from multiple providers.

\emph{(3) Hiding complexity.}
Agora hides the complexity and intricacies of assets from the consumers.
It is aware of logical equivalence of assets, i.e., assets that yield the same results (e.g.,~a nested loops join is equivalent with a hash join for equi-joins).
Implementations of logically equivalent assets can have very different properties: They may use different programming languages (e.g.,~C++ and Java), be tailored to different systems (e.g.,~Flink and Spark), be optimized for specific hardware (e.g.,~CPU and GPU), and run in a parallelized, distributed, or sequential setting.
In addition, each provider can define different pricing for her implementation.
To optimize asset execution, Agora chooses the best combination out of the available implementations based on the requirements of the incoming task or application.

\subsection{Requirements and Challenges}
To see the Agora vision become a reality, we must fulfill the following requirements:
(i)~asset sharing and discovering -- users should be able to easily provide or consume assets;
(ii)~asset privacy and security -- users must be able to set privacy and security constraints to their assets;
(iii)~asset interoperability -- users should be able to easily combine different (types of) assets;
(iv)~asset equivalence -- users should be able to achieve their desired goals without being concerned about the specifics of the underlying algorithms; and
(v)~hardware independence -- users should be able to run their assets on heterogeneous hardware seamlessly.

Ultimately, Agora aims at consolidating information from around the world and executing intelligent algorithms on top of it.
This is a formidable challenge that presents the opportunity to integrate and advance database research in many areas, from query compilation and processing to data integration and mining, while dealing with asset heterogeneity, privacy, security, heterogeneous hardware, and novel computer architectures.
Realizing this vision comes with a plethora of research questions, such as:
{\em How can we specify highly heterogeneous assets in a unified way?}
{\em Can we automatically generate such a specification?}
{\em How can we discover and potentially compose highly heterogeneous assets to satisfy a consumer's request?}
{\em What is the right pricing model for each type of asset?}
{\em How can we guarantee that for a combination of assets every contributor gets paid?}
{\em How can we specify privacy and security constraints to assets?}
{\em Can we ensure a trusted environment for the execution of assets having such constraints?}
{\em Can we enable assets to run on any computing resource of the asset ecosystem?}

\myparagraph{Outline}
In the remainder of the paper,
we first define an asset and introduce the different kinds of assets in Agora in Section~\ref{sec:preliminaries}.
Next, in Section~\ref{sec:architecture}, we present the architecture of and vision behind Agora.
In Section~\ref{sec:challenges}, we point out the research challenges and outline possible solutions.
We discuss related work in Section~\ref{sec:relwork} and conclude in Section~\ref{sec:conclusion}.

  \section{Assets in Agora}\label{sec:preliminaries}
  Before discussing the internals of Agora, let us first define assets as {\em any data-related unit of production that allows users to exploit the value of data}. We identify six major categories of assets: {\em data sources}, {\em algorithms}, {\em pipelines}, {\em systems}, {\em storage and compute resources}, and {\em applications}. In the following, we explain them as well as point out the providers' and users' incentives in each one of them:
\vspace{-2mm}
\begin{packed_enum}
  \item \textbf{Data sources.}
  These include raw data (e.g.,~relational data or graph data) as well as enriched or curated data (e.g.,~knowledge graphs and ontologies).
  In addition, data may be provided as data-at-rest (batch data) or data-in-motion (streams).
  Agora provides the platform for specialized providers that offer high-quality data.
  Such data providers can bring their data to the market and benefit from respective revenues.
  Data users benefit from the available diverse, high-quality data.

  \item \textbf{Algorithms.}
  Efficient algorithm implementations are core building blocks in data-driven applications provided by developers.
  An algorithm implementation can be part of a processing pipeline, system, or software tool.
  Typical examples include database operators, indices building, feature extraction, and ML model training.
  Agora eases code reuse as it enables secondary usage of implementations.
  For example, the databases community presents several new join algorithms at their leading conferences every year.
  However, only few of the presented algorithms see a wide-spread adoption mainly because it is hard for developers to sell/put their algorithms in the market.
  Agora enables a plug-and-play solution: any developer can offer a new join algorithm that is logically equivalent to an existing one, but more resource efficient or tailored for a specific hardware or system.

  \item \textbf{Pipelines.}
  Pipelines are a sequence of data sources and algorithms that manipulate data towards a single goal.
  The value of a pipeline lies in a ready-to-use combination of such assets.
  For example, a pipeline can combine data cleaning, feature extraction, and classification algorithms to transform raw data into labeled events.
  Setting up a pipeline of compatible algorithms is often a challenging task.
  Thus, it is attractive
  to acquire a ready-to-use pipeline, which was already tested in practice and received positive user rating, instead of implementing a new pipeline from scratch.

  \item \textbf{Systems.}
  Typical systems are relational databases, streaming engines, and ML systems.
  Each system may be proprietary or open-source.
  With Agora, users get access to different systems and can access them through one federated platform.
  This allows for testing different systems and combinations with real workloads before making a decision for production use.
  Moreover, users will find support and operation services for each system.
  System providers can offer their systems to a large number of customers without the need for individual license negotiations.
This makes it easier to bring new systems to the market and to attract users to use a system that is optimized for their workloads.

  \item \textbf{Storage and compute.}
  Agora accommodates storage and compute nodes, which can be offered by cloud providers, organizations, or individuals.
  Compute nodes can be virtual machines or dedicated servers.
  Storage resources can be main memory, disks, or network-attached storage.
  As there are diverse providers, users gain access to diverse servers with diverse hardware, can test different setups, and find the optimal environment for their application.
  In this way users avoid lock-in effects to a particular cloud provider because they can easily switch between compute nodes.
  Users can also benefit from accessing spare resources in a data center that is close to their customers or sensors.

  \item \textbf{Applications.}
  An application consists of systems, pipe\-lines, algorithms, and, optionally, data sources and storage/compute nodes to offer a complete ready-to-use solution.
  The components that constitute the application can be assets from the ecosystem or private resources.
  Application providers benefit from a platform on which they can offer applications to users similar to an app-store for smartphones.
  Application providers can develop and improve their applications using assets that are available in Agora.
  For example, one can offer a web shop as an application which integrates a pipeline for article recommendations.
\end{packed_enum}

  \section{Agora Architecture}\label{sec:architecture}
\begin{figure}[t!]
  \centering
  \includegraphics[width=\linewidth]{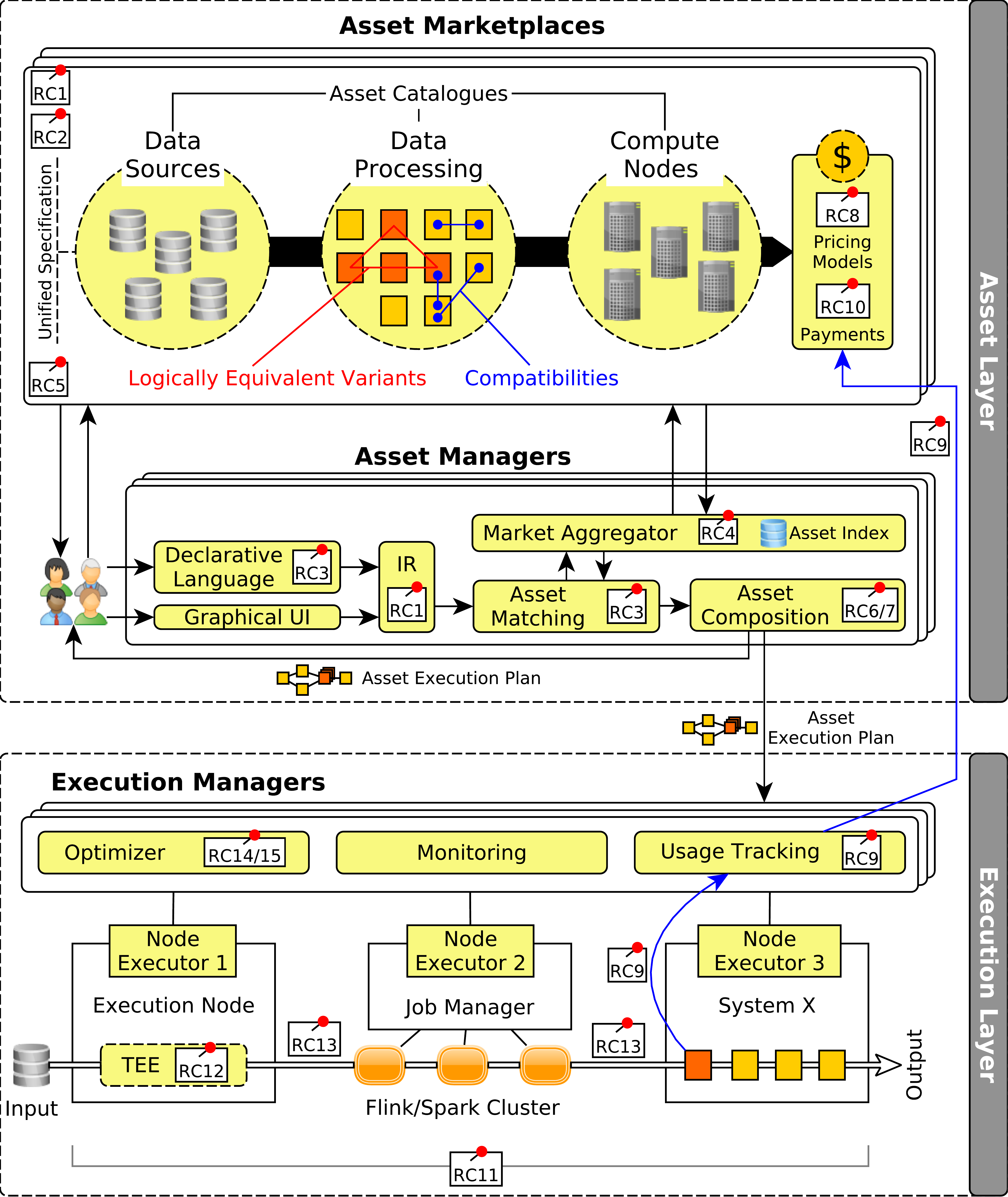}
  \caption{An overview of the architecture of Agora with 15 selected \textit{Research Challenges (RCs)}.}
  \label{fig:new-arch}
\end{figure}

  Agora builds around assets and consists of two  layers: the {\em Asset Layer} and the {\em Execution Layer}.
A major strength of Agora is its seamless connection between these two layers.
It goes beyond stand-alone marketplaces, stand-alone execution engines, and cloud services with the goal of facilitating the use of DS tools for a broader group of users.
Figure~\ref{fig:new-arch} illustrates the architecture of Agora.

The \textbf{asset layer} constitutes an ``intelligent" ecosystem of multiple {\em asset marketplaces} and enables not only offering and finding assets but also composing them in a smart way via {\em asset managers}.
Recall our motivation example described in Section~\ref{section:motivation}.
Bob, who is searching for a dataset, has the choice of going directly to his favorite marketplace or to an asset manager to find his desired dataset.
In the former case, he either browses the marketplace or uses keywords to search within it.
In the latter case, he simply specifies his request in a declarative manner and the asset manager is responsible to respond by accessing multiple marketplaces.

The \textbf{execution layer} optimizes and runs asset execution plans via {\em execution managers} and {\em node executors}.
For instance, Charlie, in our running example, finds his pipeline via an asset manager and decides to execute it in Agora.
For this reason, the asset manager translates the pipeline into an execution plan together with its equivalent assets, which are logically equivalent variants satisfying the same request.
Logically equivalent variants can be different physical implementations of the same logical operator, alternative compute nodes with similar properties, or alternative data sources, such as weather data from different providers.
Next, the asset manager passes the execution plan to an execution manager, which is responsible to optimize the plan and find the best possible equivalent pipeline asset that respects Charlie's budget.
The execution manager accesses processing nodes through a {\em node executor}, which is a standardized component to interface arbitrary execution environments with execution managers.
For example, NodeExecutor 1 in Figure~\ref{fig:new-arch} provides access to a Trusted Execution Environment (TEE), such as an Intel SGX Enclave~\cite{costan2016intel},
and NodeExecutor 2 provides access to a Flink~\cite{carbone2015apache} or Spark~\cite{zaharia2016apache} Job Manager to run Flink and Spark Jobs on a cluster.
It is worth noting that all components of Agora (asset marketplace, asset manager, execution manager, and node executor) are assets themselves.
A consumer/provider can offer her own implementation for any of these components and charge consumers for its use.
Consumers can choose between concrete implementations provided by different users.
We believe that this flexibility leads to a competition for providing the best possible Agora components, e.g., for providing the execution manager with the best optimizer.

In the following, we discuss the details of the two layers and point out 15 research challenges (RCs), which we further elaborate in Section~\ref{sec:challenges}.

\subsection{Asset Layer}\label{sec:asset_layer}
Agora's asset layer consists of an ecosystem of \textit{asset marketplaces}, which allow providers to share their assets, and \textit{asset managers}, which allow consumers to easily use assets across multiple marketplaces.

Each \textbf{asset marketplace} contains catalogues that keep track of the available assets and their properties.
To make this possible, Agora unifies assets under a common specification.
Only a unified specification enables easy asset discovery and composition across all the marketplaces in the ecosystem.
Providers should conform with this unified specification when they offer new assets to the marketplaces.
This can be a barrier for new asset providers.
Therefore, it is crucial that Agora provides the means for automatically generating asset specifications from more intuitive user inputs, such as query and programming languages or graphical interfaces.
Defining such a specification and determining ways for its automated extraction is challenging due to the the large heterogeneity of assets (\textit{RC1} and \textit{RC2}).

Moreover, providers might want to specify usage constraints to their assets.
For example, location requirements (e.g.,~private data may not be moved out of a country) or vendor requirements (e.g.,~my algorithm may not be used by a competitor) may be asset constraints.
Identifying the best way to describe constraints over assets is an interesting research challenge because of the asset heterogeneity and different constraint granularity (\textit{RC5}).

Providers can also define a pricing model (e.g.,~subscriptions or pay-per-use) for their assets usage (\textit{RC8}).
Ideally, Agora proposes a pricing model and a price based on monitoring the current trend of the market.
When a provider chooses a pay-per-use pricing model, Agora ensures to track the asset's usage and report usage counters back to marketplace (\textit{RC9}).
Marketplaces then perform the invoicing and initiate (micro-)payments between users (\textit{RC10}).

\textbf{Asset managers} are the entry point for users who want to declaratively: find assets across different marketplaces; combine multiple assets into execution plans; and run asset execution plans.
An asset manager provides a graphical user interface and/or a declarative language for finding and composing assets~(\textit{RC3}).
A user request is then converted to an intermediate representation (IR), which allows for matching asset specifications with user requests (\textit{RC1}).
The asset manager matches user requests to assets that are compatible with each other and satisfy the requests (\textit{RC3}).
For this, it needs to aggregate the assets of all marketplaces and build an asset index (\textit{RC4}).
Next, the asset manager composes all the relevant assets (with their equivalent assets) together so that they fulfill the request.
When composing assets, it is crucial to satisfy usage constraints of the assets (\textit{RC6} and \textit{RC7}).
As a result, the asset manager outputs an asset execution plan, which allows the execution layer to further optimize, deploy, and run the plan.

\subsection{Execution Layer}\label{sec:execution_layer}
Agora's execution layer consists of \textit{execution managers}, which receive execution plans from an asset manager, and \textit{node executors}, which allow consumers to run their assets.

An \textbf{execution manager} is a core component of the execution layer.
It is responsible for optimizing an asset execution plan, deploying it on compute nodes, and monitoring its execution.
As the plan may contain different variants of operations, the execution manager can schedule an operation of an execution plan on different execution environments (node executors).
Achieving this multi-environment execution of a plan is very challenging as the search space of all possibilities to execute a plan becomes very large (\textit{RC14} and \textit{RC15}).
The selection of existing variants and the selection of node executors goes hand-in-hand with possible algorithm adaptations, which increases the performance on a particular target system.

A \textbf{node executor} is Agora's interface component to connect arbitrary execution environments with execution managers.
For example, in Figure~\ref{fig:new-arch} the asset execution plan is deployed to three node executors with different characteristics:
NodeExecutor~1 provides access to a trusted execution environment (TEE), which provides additional security because the owner of the node has no access to the executed source code nor the processed data (\textit{RC12});
NodeExecutor~2 provides access to a Flink or Spark Cluster; and
NodeExecutor~3 provides direct access to hardware resources on a dedicated server.
When dealing with multiple node executors, Agora provides a secure way to transfer data among nodes to validate data integrity and to pay for data that is traded as an asset.
This is hard to achieve especially when data is large or data streams have high bandwidth~(\textit{RC13}).

It is worth noting that both node executors and execution managers are responsible for tracking the usage of assets, which is crucial to ensure fair payments.
This is a challenging task because it also assumes tracking fine-granular operations in a composition of assets~(\textit{RC9}).
Agora adopts certificates to ensure transparency and trust between consumers and providers.
For example, one can certify the physical location of a node, security standards, compliance with asset usage tracking, or energy efficiency.
The main challenge remains in the standardization of certificates and assets requirements (\textit{RC11}).

  \section{Research Directions}\label{sec:challenges}
  We now elaborate on the 15 main research challenges that we believe are crucial to address in order to implement Agora.
Most of the challenges stem from the heterogeneity of assets and the open ecosystem setting.
They deal with asset management (Section~\ref{sec:asset-management}), compliant  asset processing (Section~\ref{sec:complient-processing}), pricing and payments  (Section~\ref{sec:trackingandbilling}), privacy and security (Section~\ref{sec:privacy-and-aecurity}), and efficient asset execution (Section~\ref{sec:efficient-asset-execution}).
In the following, we discuss each research challenge and outline approaches to tackle them.

\subsection{Asset Management}\label{sec:asset-management}
The first step towards Agora is enabling effective and efficient \emph{asset management}: any asset-related operation, such as asset sharing, discovery, and composition.
We identify the following four main research challenges that we need to tackle to achieve this.

\challenge{Unified specification}
A major challenge for asset management is the design of a unified specification (a {\em standard}).
Such a specification will allow sharing and discovery of assets not only within a single marketplace but also among different markets.
It, thus, facilitates the usage of an asset search engine across different marketplaces.
The difficulty in devising such a standardization lies in the fact that there are different types and granularities of assets: from datasets and stream sources to complex algorithms or data management systems. The standard should take all these different types of assets into consideration while keeping as much simplicity as possible.
In addition, a single asset may not be sufficient to satisfy a consumer’s request. For this reason, the standard should enable interoperability among assets so that composite assets, i.e.,~assets formed by multiple assets, can also be shared.
To enable asset composition, such as the one required for our example in Section~\ref{section:motivation},
the specification must be flexible to consider asset combinations.
It should enable building complex pipelines and systems and at the same time be general enough to support all operations and multiple query languages.

\begin{figure}[t]
  \centering
  \includegraphics[width=\columnwidth]{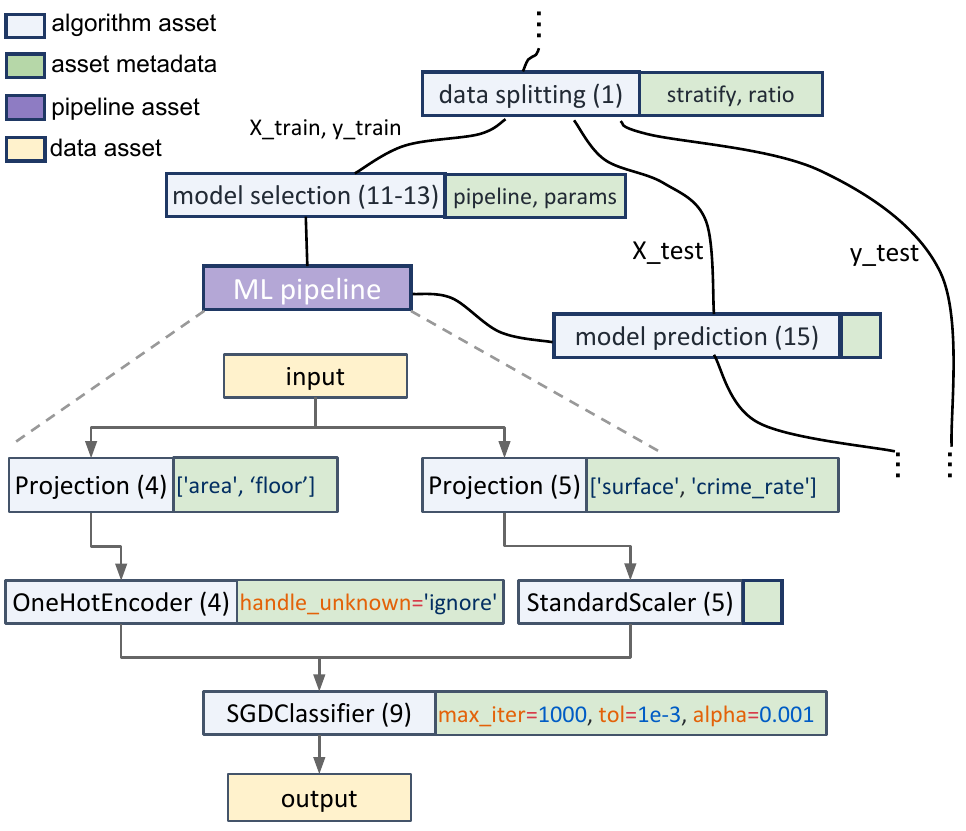}
  \caption{Automated specification generation of the asset shown in Listing~\ref{lst:python}. The numbers next to the algorithm assets point to the lines of code in Listing~\ref{lst:python} that represent that particular asset.}
  \label{fig:simple_example_ir}
\end{figure}

Our initial efforts towards a unified specification is a declarative intermediate representation of data science assets~\cite{Redyuk2019a}. To cope with the lack of higher-level declarative abstractions for end-to-end data science processes~\cite{Schelter2018}, we have defined \textit{a schema for the specification of the execution of data science pipelines} inspired by ML Schema~\cite{Publio2018} or Amazon's experiment tracker~\cite{Schelter2017}. Figure~\ref{fig:simple_example_ir} shows an example of the intermediate representation of a data science pipeline asset following this schema specification. Nodes in the graph represent high level asset categories, optionally accompanied by their metadata, and edges connect two assets by their input/output. Having such a high level representation of assets allows us to make further optimizations and find equivalences among different assets.

\begin{lstlisting}[caption=Excerpt of a data science pipeline asset expressed in Python which predicts real estate pricing., label=lst:python, float]
X_train,X_test,y_train,y_test = train_test_split(X,y,test_size=.1)

feature_transformation = ColumnTransformer(transformers=[
 ('categorical_attr', OneHotEncoder(unknown='ignore'), ['area', 'floor']),
 ('numeric_attr', StandardScaler(), ['surface', 'crime_rate'])])

pipeline = Pipeline([
 ('features', feature_transformation),
 ('learner', SGDClassifier(max_iter=1000, tol=1e-3))])

param_grid = { 'learner__alpha': [0.0001, 0.001, 0.01, 0.1] }
search = GridSearchCV(pipeline, param_grid, cv=5)
model = search.fit(X_train, y_train)

predicted = model.predict(X_test)
\end{lstlisting}

\challenge{Automated specification generation}
Providing the asset specification can be error-prone and introduce significant overhead to asset providers.
For this reason, it is necessary to provide mechanisms to generate assets specification from more intuitive user inputs.
This opens up new research directions on automated extraction of a specification from query and programming languages as well as graphical user interfaces.
Our first step towards this direction focuses on data science asset providers, i.e., data scientists, who are primarily familiar with writing Python scripts.
Agora extracts an intermediate representation from data science pipelines written in Python code with little or no involvement of the end-user~\cite{Redyuk2019a}.
Our approach allows for simple and straightforward use of the asset layer, yet integrates it with a powerful tool for search and sharing, potentially across languages and domains.
Automation of schema extraction is based on static code analysis~\cite{Namaki2020} and its semantic enrichment~\cite{Patterson2018}.
Listing~\ref{lst:python} shows a real estate predictor asset, while Figure~\ref{fig:simple_example_ir} shows the automatically generated intermediate representation of this asset.
To achieve this automated extraction we maintain a simple knowledge base consisting of data science sub-processes (e.g., normalization).
We then map object signatures that the programming language (and its ecosystem) supports to the categories found in the knowledge base.
For example, \texttt{Python}'s \texttt{sklearn.preprocessing.OneHotEncoder} class signature maps to the `data-preprocessing-transformation' category, that instructs the system what meta information to extract and how.
We plan to further investigate this direction and attempt exploiting pattern mining solutions as a potential replacement of manually curated knowledge bases.

\challenge{Matchmaking}
The asset layer via its asset search engine (as well as a single marketplace) should be able to effectively and efficiently identify all assets related to a given consumer's request.
To achieve this the marketplace should provide the users with a declarative query language or graphical user interface that allows them to discover assets with the desired characteristics.
Using the graphical user interface lay users can browse assets or use a keyword search, while more advanced users should be able to use the declarative query language to quickly describe the assets they want.
Devising a declarative language which can express requests about different types and granularities of assets is a challenging task.
In addition, identifying the most suitable approach for matching a query with the available assets is not straightforward.
To solve this challenge we are looking into the direction of matchmaking and recommendation, which has been used recently in multi-sided marketplaces~\cite{recsys19-tutorial}.
The difference with traditional recommendation systems is that in the case of marketplaces, such as the asset layer we envision in Agora, there is a multi-objective optimization problem that needs to be taken into consideration: increasing both provider and consumer satisfaction.
We also plan to combine recommendation systems with the solutions that focus on satisfaction-based~\cite{sqlb,sblq-journal,sbqa} and economic-based query processing~\cite{mariposa}.

\challenge{Market aggregator}
The Agora ecosystem is composed of multiple asset marketplaces.
It is thus important for the asset manager to be aware of the different marketplaces and their assets through a market aggregator.
The challenge we have to face when building the market aggregator is twofold: (i)~indexing available assets in an efficient and scalable way despite their number and diversity,  and (ii)~finding equivalences among assets.
Although the asset specification facilitates the comparison between two assets, it is still not straightforward how exact or approximate equivalences can be found.
We plan to incorporate techniques from source code search engines~\cite{sourcecode-searchengine} and program translation~\cite{program-translation}, traditionally used to migrate code from one language to another, to tackle these two challenges.

\subsection{Compliant Asset Processing}\label{sec:complient-processing}
In such an open asset-centric ecosystem as Agora, it is important to allow providers to provide constraints to their assets.
A provider might not want her asset to be processed in unintended ways and therefore may specify usage policies that the asset consumer should comply to. For this reason, asset processing (i.e., satisfying a user's request) faces unique challenges due to asset constraints and legal requirements.
For example, a usage policy may prohibit overlaying (joining) the provided data with any other data~\cite{navteq} or may disallow aggregation with other providers~\cite{yelp}.
Moreover, combining geo-distributed assets may involve transfer or shipping of assets across borders.
As a result, asset processing must comply to regulations (such as GDPR~\cite{gdpr} or CCPA~\cite{ccpa}) that prohibit the use or flow of assets across geographical borders or certain sites.
For example, processing data generated by autonomous cars in three different geographies, such as Europe, North America, and Asia, may face different regulatory constraints:
There may be legal requirements that only aggregated or anonymized data may be shipped from Europe and no data whatsoever may be shipped out of Asia.
This opens up a completely new dimension of \emph{compliant query (``asset'') processing} that entails the following two research challenges.
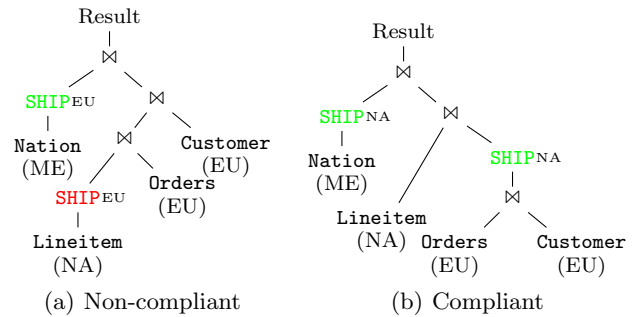
\begin{figure}
  \tikzset{
    ship/.style={draw,circle, minimum size = 2mm, inner sep=0pt},
    traits/.style={draw,rectangle, minimum size = 1mm, inner sep=0pt},
    black/.style={fill=black, text=white},
  }

  \small
  \mbox{
  \subfigure[Non-compliant]{
    \label{subfig:non-compliant}
    \begin{tikzpicture}
    \node[] (Q) {Result};
    \node[below=2mm of Q] (j1) {$\Join$};
    \node[below left=2mm and 2mm of j1, label={[label distance=-2mm]0:{\tiny EU}}] (ship1) {\textcolor{green}{\texttt{SHIP}}};
    \node[below=2mm of ship1] (n) {\shortstack{\texttt{Nation}\\(ME)}};
    \node[below right=2mm and 2mm of j1] (j2) {$\Join$};
    \node[below left=2mm and 0mm of j2] (j3) {$\Join$};
    \node[below right=2mm and 0mm of j2] (c) {\shortstack{\texttt{Customer}\\(EU)}};
    \node[below left=4mm and 0mm of j3,label={[label distance=-2mm]0:{\tiny EU}}] (ship2) {\textcolor{red}{\texttt{SHIP}}};
    \node[below right=2mm and 0mm of j3] (o) {\shortstack{\texttt{Orders}\\(EU)}};
    \node[below=2mm of ship2] (l) {\shortstack{\texttt{Lineitem}\\(NA)}};

    \draw[-]
    (Q) -- (j1) -- (ship1) -- (n)
    (j1) -- (j2) -- (j3) -- (ship2) -- (l)
    (j2) -- (c)
    (j3) -- (o)
    ;
    \end{tikzpicture}
  }
  \subfigure[Compliant]{
    \label{subfig:compliant}
    \begin{tikzpicture}
    \node[] (Q) {Result};
    \node[below=2mm of Q] (j1) {$\Join$};
    \node[below left=2mm and 2mm of j1, label={[label distance=-2mm]0:{\tiny NA}}] (ship1) {\textcolor{green}{\texttt{SHIP}}};
    \node[below=2mm of ship1] (n) {\shortstack{\texttt{Nation}\\(ME)}};
    \node[below right=2mm and 2mm of j1] (j2) {$\Join$};
    \node[below left=10mm and 0mm of j2] (l) {\shortstack{\texttt{Lineitem}\\(NA)}};
    \node[below right=2mm and 2mm of j2, label={[label distance=-2mm]0:{\tiny NA}}] (ship2) {\textcolor{green}{\texttt{SHIP}}};
    \node[below=2mm of ship2] (j3) {$\Join$};
    \node[below left=2mm and 0mm of j3] (o) {\shortstack{\texttt{Orders}\\(EU)}};
    \node[below right=2mm and 0mm of j3] (c) {\shortstack{\texttt{Customer}\\(EU)}};

    \draw[-]
    (Q) -- (j1) -- (ship1) -- (n)
    (j1) -- (j2) -- (ship2) -- (j3) -- (c)
    (j2) -- (l)
    (j3) -- (o)
    ;
    \end{tikzpicture}
  }}
  \caption{Excerpt of distributed query plans for TPC-H Query 10. The leaf nodes denote base tables located in Middle East (ME), North America (NA), and Europe (EU).}
  \label{fig:queryplans}
\end{figure}

\challenge{Constraint specification}
The first challenge to overcome is determining how to specify asset constraints declaratively.
Doing so is important for easing the specification of constraints.
However, it is challenging not only because of the asset heterogeneity but also because of the different constraint granularities.
For example, a constraint might apply to an entire asset, parts of it, or even to information derived from it.

\challenge{Constraint satisfaction}
The second challenge is to find efficient ways to process queries in a manner compliant with respect to asset constraints.
In our early efforts towards realizing Agora we provide support for
{\em compliant geo-distributed} query processing.
Our initial implementation allows expressing constraints on shipping data across geographical borders using our \emph{extended}-SQL statements.
Its query optimizer aims at finding distributed query execution plans that are compliant with respect to shipping of intermediate data between compute sites.

To illustrate query plans produced by a compliant query optimizer, assume TPC-H query $Q_{10}$ in a setting where data is geo-distributed: the base tables are geo-distributed across the Middle East (ME), North America (NA), and Europe (EU).
Also, we set one constraint stating that no data from NA can be shipped to EU.
Figure~\ref{fig:queryplans} shows excerpts of the query plans produced by a traditional query optimizer (Figure~\ref{subfig:non-compliant}) and our optimizer (Figure~\ref{subfig:compliant}).
The query plan on the left is not compliant because it disregards constraints on shipping parts of the Lineitem table to Europe.
\begin{wrapfigure}{rh}{0.24\textwidth}
       \centering\includegraphics[scale=0.3]{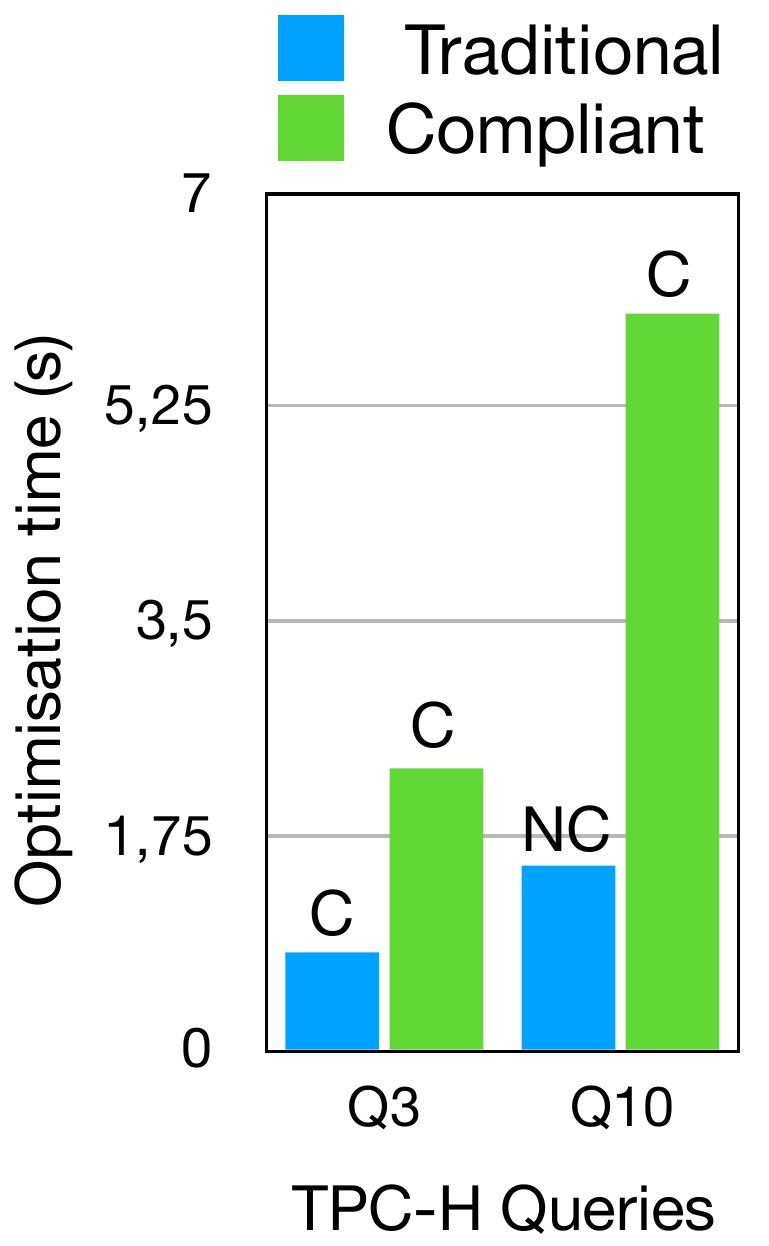}
\caption{Traditional vs compliant query (asset) processing; Letters on top of
  bars denote if the plan was compliant (C) or non-compliant (NC).}
\label{figure:compliant}
\end{wrapfigure}

Figure~\ref{figure:compliant} shows the query optimization time for that query along with query TPC-H $Q_{3}$.
Both queries involve joining data from different geographical sites, with a number of constraints on data movement across the different geographical sites.
We observe that traditional query processing is not suitable for such settings
as they simply disregard the data movement constraints: they can indeed
produce non-compliant query execution plans (denoted by NC), such as the one
for $Q_{10}$, whereas our approach always produces a compliant plan (denoted by
C) if it exists.
This shows that traditional query processing techniques are unsuitable for dealing with asset constraints.
Still, is it possible to match (or be as close as possible to) the performance of traditional query processing?
This is a major challenge we are seeking to tackle in Agora.

\challenge{Capturing asset provenance} Another important aspect when combining
and sharing assets is to be able to audit compliance with respect to data
usage and its sharing policies. To this end, we also need provenance
capturing technology in an asset-centric marketplace. While work on using
provenance to audit compliance (e.g.,~\cite{Chothia2016, datalawyer}) has
received much traction, their applicability is limited to homogeneous
execution environments or to special data processing facility.
To support auditing in Agora, we need novel solutions that can capture provenance
in a heterogeneous execution environment. In particular, we need provenance
models that can capture relationships in composite assets, deal with
diverse data models of assets, and cope with large amounts of provenance data.

\subsection{Pricing and Payments}\label{sec:trackingandbilling}
In contrast to existing marketplaces for data or algorithms and to existing cloud providers, Agora leverages a more flexible and extremely diverse combination of assets.
This makes it challenging to track each stakeholder's contributions and consumptions and to organize the respective invoice and payment processes. In this section, we discuss research directions with respect to pricing and billing in Agora.

\challenge{Pricing models}
Our ecosystem should allow providers to define prices of their assets based on different pricing models.
Ideally, the system should also propose a price based on a continuous market monitoring.
Ideas from query-based pricing~\cite{DBLP:conf/icde/DashKA09,Koutris-pods12} and economic models for the cloud~\cite{mariposa} can be the foundation, but have to be extended to fit a more general data ecosystem.
We plan to support different pricing models.
In software licensing, there are three common and fundamentally different pricing models: pay-once, subscription, and pay-per-use.
With \textit{pay-once}, a user buys a license once and can use the licensed software forever.
\textit{Subscription} models are similar to the pay-once model, with the difference that licenses may expire and have to be renewed.
The \textit{pay-per-use} model is common for cloud services where users pay every time they use a service or call a function (e.g.,~the Google Speech API, the Twitter API, and AWS Lambda functions).
A provider could adopt any of these models.
For instance, pay-per-use can be used for algorithms (e.g.,~pay \$1 per thousand calls) and for compute resources (e.g.,~pay \$5 per hour).
While a pay-per-use model seems to be the fairest solution, it is challenging to realize pay-per-use in a processing pipeline that consists of many different assets including algorithms, code, and compute resources.
In the following, we layout the challenges related to usage tracking and micro-payments as well as outline a solution for each of them.

\begin{figure}[t]
  \centering
  \centering\includegraphics[scale=0.3]{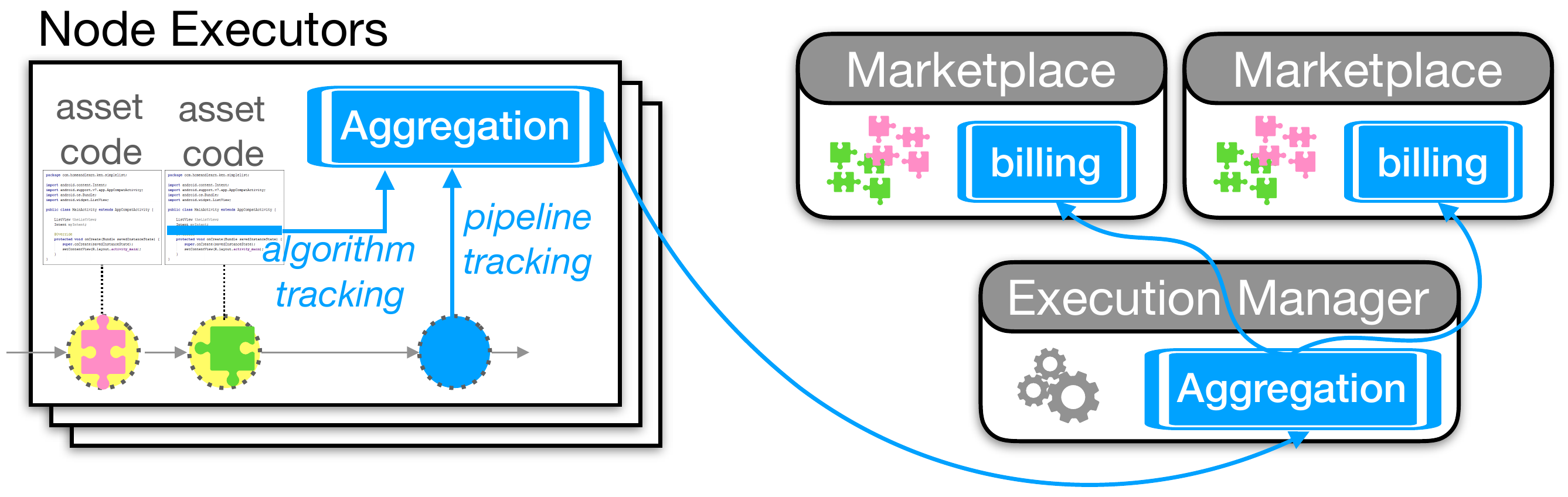}
\vspace{1mm}
  \caption{Asset usage tracking and billing.}
  \label{fig:tracking-and-billing}
\vspace{1mm}
\end{figure}

\challenge{Asset usage tracking}
To ensure fair asset payments, the execution manager should be able to track the usage of the assets.
However, tracking fine-granular operations in a set of assets (e.g.,~in a pipeline), which may run in parallel, is not an easy task.
It requires not only an aggregation component but it also depends on the trustworthiness of the nodes that report the usage tracking.
In Figure~\ref{fig:tracking-and-billing}, we depict a possible mechanism for usage tracking.
This mechanism provides a common API that allows for calling a tracking function from the asset source code (to track the use of assets) or as an operator (to track the use of pipelines).
Alternatively, one could also track the amount of processed data as part of our secure transmission process (see Section~\ref{sec:privacy-and-aecurity}).
Because usage tracking functions are called many times (e.g.,~per processed tuple), an aggregation component is required to propagate aggregated usage counters (e.g.,~once per minute) instead of individual function calls.
We plan to base this aggregation component on our previous work on efficiently aggregating data streams~\cite{carbone2016cutty, traub2018scotty, traub2019efficient} as well as on related work that enables distributed (pre-) aggregation~\cite{benson2020Disco, madden2002tag}.
Still, such a usage tracking mechanism does work only if compute nodes honestly report usages counters.
We, thus, allow for restricting the execution of operators and pipelines to specific nodes, which fulfill certification requirements (see Section~\ref{sec:privacy-and-aecurity}).

\challenge{Payments}
Ensuring a safe way for providers to charge and consumers to pay the use of assets is crucial for the ecosystem health.
Ideally, a payment process would be distributed such that each component can receive micro-payments and forward parts of these payments to sub-components.
For example, an execution manager may charge \$1 to process a MB of data, but has to share that money with asset providers.
Note that composite-asset providers have to split their share again, to pay the individual asset providers that are part of the pipeline.
Recently, blockchain-based techniques~\cite{lundqvist2017thing, xu2016blockchain} as well as blockchain-alternatives such as IOTA~\cite{popov2016tangle} have been proposed to support such micro-payments.
However, all these techniques have been criticized for either limited scalability, transaction-fees, proof-of-work requirements, security issues, missing final settlement of transactions, or authority centralization.
Morevoer, given the diversity of cryptocurrencies and their underlying technologies~\cite{corbet2019cryptocurrencies}, it is impossible to select a single best payment system.
Therefore, we aim at integrating an abstraction layer to make Agora agnostic to the details of the payment method used between users.
Agora will provide a reference implementation for the most common payment methods and users may implement additional options:
users will have to implement the logic for executing a payment, including a notification about the completion of a transaction;
Agora will trigger transactions and confirm completed transactions based on the users' implementation.

\subsection{Privacy and Security}\label{sec:privacy-and-aecurity}
Another major concern in an open ecosystem is privacy and security.
Agora needs to ensure privacy and security when processing assets as well as secure, private, and scalable data transfer among users and processing nodes.
We describe both aspects and present respective research directions in the following.

In Agora, users may decide to run their assets on processing nodes operated by a diversity of providers.
As these providers have physical access to their processing nodes, they potentially gain access to the code of assets that runs on their nodes and the data these nodes process.
Both data and code of assets should be protected against unauthorized access and manipulation to ensure privacy and to prevent attacks aiming at manipulating results.
We investigate three approaches that complement each other:
establishing {\em trust certificates}, using {\em trusted execution environments}, and ensuring {\em secure data transfer}.

\challenge{Establishing trust certificates}
Certifications are a common way to establish trust between cloud providers and users~\cite{sunyaev2013cloud}.
However, existing certifications for cloud providers assume a single provider (e.g., AWS, Microsoft Azure, or IBM Bluemix) to serve a very large number of users.
Thus, the certification process can be complex and users are able to check certificates manually for the (only) one provider they use.
Agora aims at drastically increasing flexibility for asset creation and execution.
Consequently, the main challenge resides in the standardization of certificates and asset requirements.
Our goal is to enable the execution manager of Agora to automatically match assets with compute and storage resources.
To this end, our key idea is to democratize the certification of properties, such as security standards and the locations of nodes.
Everyone can become a certification authority and decide which authorities to trust.
For example, the EU could certify that a compute node is located in the EU and therefore become a certification authority.
The execution constraints of an asset (or asset execution plan)
then include a set of required certificates connected with trusted authorities for each type of certificate.
Technically, we plan to use the TLS handshake protocol~\cite{morrissey2010tls} as solution for authenticating compute node properties.
In contrast to common TLS in the world wide web, each compute node in Agora may hold a plethora of certificates issued by diverse certification authorities.
The execution manager then validates that all required certificates are present at a compute node before assigning an asset to that particular node.

\challenge{Trusted execution environments}
A Trusted Execution Environment (TEE) provides a solution for secure computation, which does not require to trust the owner of a compute node.
Thus, TEE-based solutions go beyond certification-based solutions to protect assets code and data, which are particularly critical for security.
We especially consider TEEs that enable remote execution, such as ARM TrustZone~\cite{ngabonziza2016trustzone} and Intel Software Guard
Extension (SGX)~\cite{costan2016intel}.
The key idea is that processor vendors provide a secure execution environment within their processors.
The processor ensures the integrity of the executed code with a remote attestation, which prevents code manipulations~\cite{wang2017enabling}.
All data enters the secure environment encrypted, and is decrypted only within the processor.
The processor also encrypts all outputs before they leave the secure environment.
Thus, the owner of a compute node cannot see or manipulate any asset data or code that runs inside the TEE, i.e.,~within the processor.
In the past, it was difficult to engineer applications for TEEs, which has also lead to security vulnerabilities~\cite{he2018sgxlinger, tang2017clkscrew}.
Nowadays, open source frameworks, such as Asylo~\cite{porter2018introducing} and Keystone~\cite{lee2019keystone}, ease the development of assets that run in TEEs.
This makes it feasible to leverage TEEs in the context of distributed data processing.
Agora will support TEEs to improve security in general and to enable secure data processing even on uncertified nodes.
Thereby, existing works on TEE-secured databases~\cite{priebe2018enclavedb, vinayagamurthy2019stealthdb, zheng2017opaque} and stream processing systems~\cite{havet2017securestreams, park2019streambox, thoma2019behind} are an important first step, but need to be extended to be (i) scalable, (ii) generally applicable, and (iii) easy to use in the context of an asset-based ecosystem such as Agora.

\begin{figure}[t]
  \centering
  \includegraphics[scale=0.35]{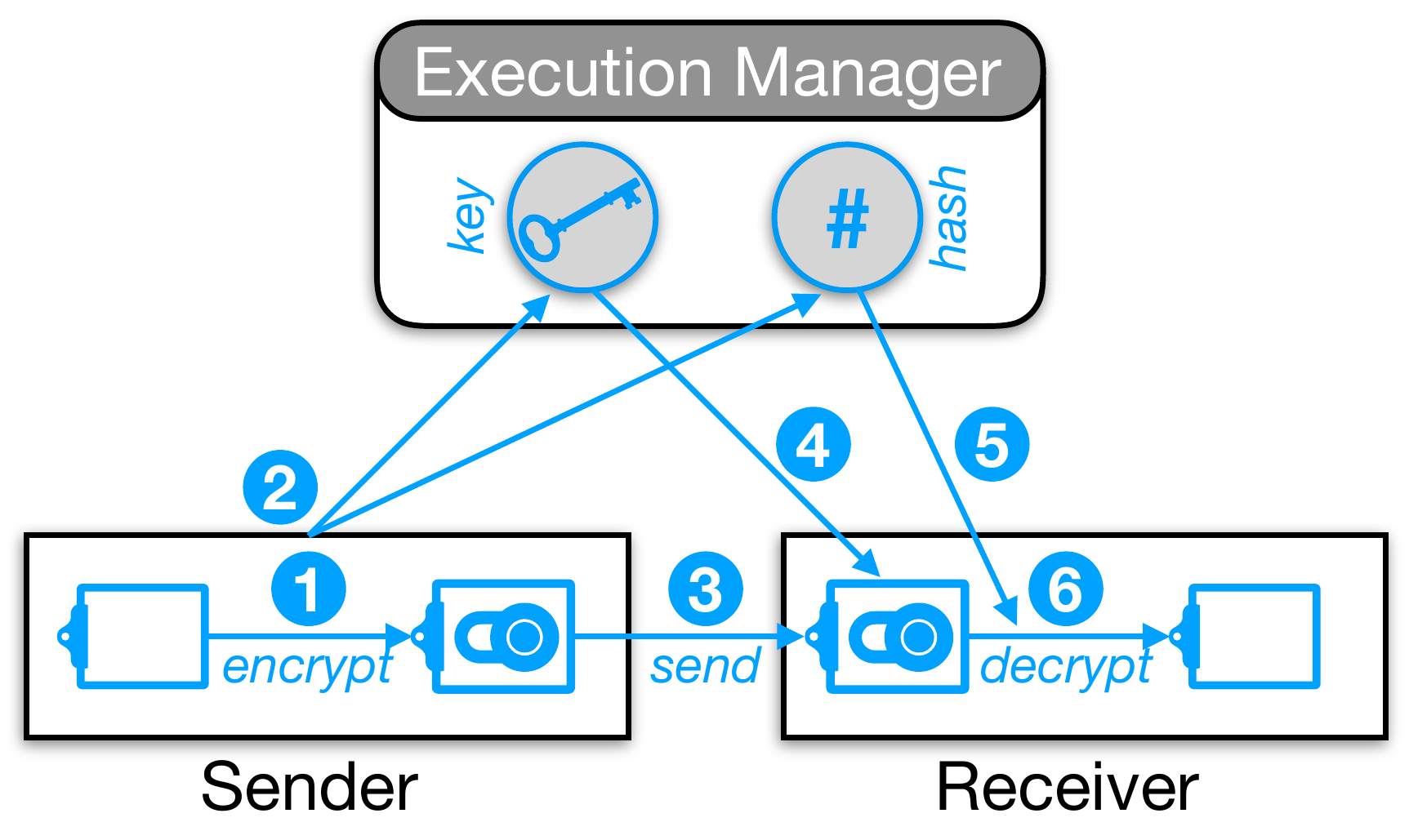}
  \caption{Secure Data Transfer and Escrow between two node executors.}
  \label{fig:secure-data-transfer}
\end{figure}

\challenge{Secure data transfer}
It is important that users can exchange data among them in a {\em secure} way within Agora.
In this context, `secure' means that
\begin{enumerate*}[label=(\roman*)]
  \item all data transmission is encrypted to prevent unauthorized access,
  \item the integrity of the data is guaranteed and can be validated by receivers, and
  \item sender and receiver can use an escrow service to secure data trading.
\end{enumerate*}
One of the challenges is that data can be arbitrarily large and data streams often have high bandwidths.
Thus, senders need to send data directly to receivers and the execution manager should act as a coordinator.
We outline our solution for secure data exchange in Figure~\ref{fig:secure-data-transfer}.
The execution manager acts as a mediator to pass the hash value and key of the encrypted data from the sender to the receiver.
Thus, the execution manager works without storing or transmitting the data itself, which prevents it from becoming a honeypot of data for potential attackers.
The execution manager releases the key if and only if the receiver issues the payment for the received data.
The receiver will only issue the payment once it confirmed the data integrity using the provided hash value.
Existing key escrow encryption services~\cite{denning1996taxonomy} can serve as a blueprint for our ecosystem.
However, we need to adapt these techniques to support assets requiring stream processing and intermediate transmissions within assets.
We want to design a scalable and light-weight escrow process, which can be performed even for small chunks of data (e.g., network packages).
This process has to combine fast micro-payments (discussed in Section~\ref{sec:trackingandbilling}) with a scalable implementation of the coordination component in the execution manager.

\subsection{Efficient Asset Execution}\label{sec:efficient-asset-execution}
Given the high diversity of assets in Agora, it is crucial to also provide a diverse execution environment in order to obtain maximum performance.
Following the one-size-does-not-fit-all dictum, a plethora of specialized systems have emerged since almost two decades ago.
There are reportedly over $200$ different platforms only under the umbrella of NoSQL~\cite{dbengines}.
Each excels in specific aspects, e.g.,~Spark is optimized for batch processing (requiring full scans) and a database is very efficient for point queries (requiring index access), leading to works using multiple systems~\cite{systemml, musketeer,ml4all,comparison,thecase}.
At the same time, processor vendors have turned to specialization and acceleration, i.e., building processors that are optimized for a specific use case~\cite{Borkar_2011_The-Future}, such as GPUs and FPGAs.
Broadly speaking, GPUs are optimized for highly parallel throughput applications~\cite{Lindholm_2008_NVIDIA}, whereas CPUs are optimized for single thread performance~\cite{Borkar_2011_The-Future}.
FPGAs, in turn, enable the design of custom hardware solutions to meet high demands on latency and throughput and hence are also increasingly being used to accelerate some data processing tasks~\cite{Teubner_2013_Data,Fang_2019_In-memory}.

In this highly heterogeneous computing landscape, it is crucial that Agora fully leverages the capabilities of each data processing platform (databases, dataflow-based processing systems, stream processing systems, etc.) and computing device (CPU, GPU, or/and FPGA) to get the maximum performance benefits out of them.
However, fully leveraging this heterogeneous computing landscape is challenging for several reasons that we explain in the following.

\challenge{Heterogeneous asset deployment}
Agora can determine the deployment environment, i.e.,~the processing system for deploying each asset.
\begin{wrapfigure}{rh}{0.22\textwidth}
  \centering
  \includegraphics[scale=0.32]{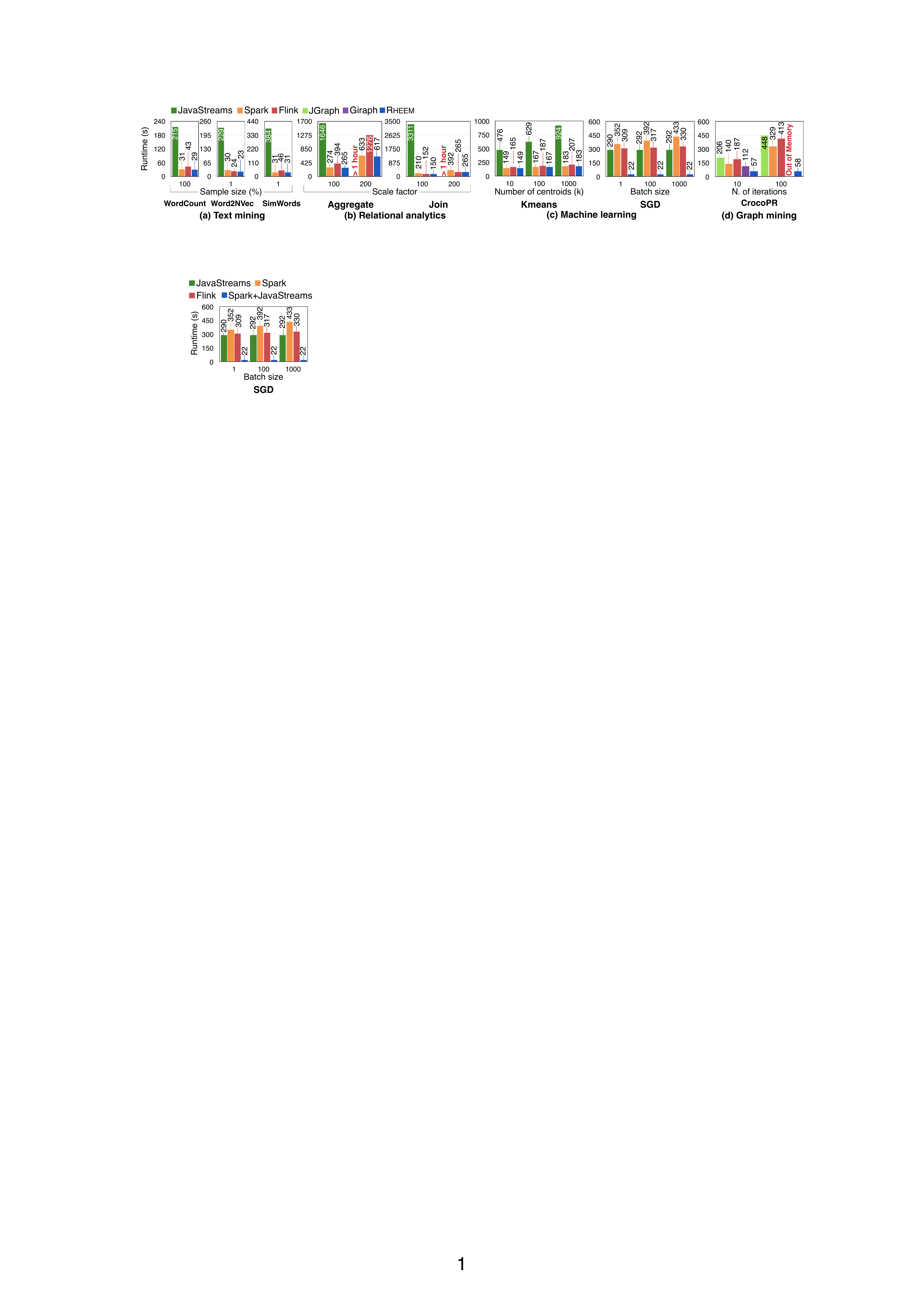}
  \caption{Benefits of heterogeneous asset deployment for SGD.}
  \label{figure:sgd}
\end{wrapfigure}
For example, if the asset is a stream processing algorithm, Agora might decide to run it on Flink~\cite{carbone2015apache}, while if it is a reinforcement learning algorithm, it may decide to run it on Ray~\cite{ray}.
Identifying the type of assets and where they should be executed is a very challenging task.
We already did the first step towards this direction with Rheem~\cite{rheem}.
We have shown that using multiple data processing platforms significantly decreases the execution time of a single processing task.
For instance, Figure~\ref{figure:sgd} shows the runtime of a classification training asset using stochastic gradient descent (SGD) for different batch sizes and for the HIGGS dataset as input.
We observe that enabling heterogeneous asset deployment (Spark and JavaStreams for the example) can significantly increase performance (more than one order of magnitude faster than using Spark, Flink, or JavaStreams only for the example).
We also have shown such performance benefits for a large variety of other tasks~\cite{rheem,ml4all}.
Thus, the consumers of Agora can benefit from such performance increase without any knowledge about the deployment itself.
Although Rheem is one of Agora's ingredients, considering highly diverse assets is still an open research problem.

\challenge{Heterogeneous asset execution}
In addition to determining which processing system to execute an asset, Agora also determines how to allocate the asset to compute resources.
Given a number of processor-specific algorithm implementations, it has to decide on which processor to execute every single asset.
However, achieving this in an automatic way is challenging for several reasons.
To statically schedule assets, we have to specify the computational requirements of an algorithm as metadata and match them with the computational capabilities of hardware providers.
Scheduling assets dynamically at runtime requires cost models and output cardinality estimates that capture algorithm behavior on heterogeneous computing resources.
Although such cost models~\cite{He_2009_Relational,He_2013_Revisiting} and cardinality estimates~\cite{Gregg_2011_Where,Yuan_2013_The-Yin-and-Yang} exist for specific applications, Agora requires more generic models to reflect the asset diversity.
A promising approach is to synthesize complex algorithms from basic data layout design choices and data access primitives, which one can quickly benchmark on different processors~\cite{Idreos_2018_The-Data}.
This approach has been demonstrated only on CPUs.
We will extend these basic building blocks to capture the specific properties of heterogeneous computing resources.
Still, Agora must adapt algorithms to the specific processor they run on to exploit the full potential of heterogeneous computing resources.
For this, we must automatically generate such processor-specific algorithm implementations.
Our previous work~\cite{Bress_2018_Generating,Rosenfeld_2019_Performance,Rosenfeld_2015_The-Operator} demonstrates that this is indeed feasible:
Data processing systems can learn processor-specific implementations during installation or at runtime.

\vspace{2mm}\noindent
The abovementioned research challenges present the opportunity to integrate and advance database research in many sub-fields, from query compilation and processing to information integration to data mining, while dealing with privacy, security and billing as well as with heterogeneous hardware and other novel computer architectures.

  \section{Related Work}\label{sec:relwork}
  Most advanced DS systems require huge amounts of data, cutting edge data science innovations, and powerful computational infrastructure.
Agora aims to connect providers and users of these key assets in an open ecosystem.
In contrast, recent works such as OpenAI~\cite{openai}, Ocean Protocol~\cite{oceanprotocol}, ML Bazaar~\cite{mlBazaar}, Enigma~\cite{enigma}, Datum~\cite{datum}, and Nebula~\cite{nebula} tackle only parts of the solution provided by Agora.
For example, OpenAI~\cite{openai} is the first non-profit research initiative promoting ``openness'' in AI.
This organization aims at ensuring that AI benefits touch all of humanity.
However, it primarily builds custom solutions and shares them via free software for training, benchmarking, and experimenting.
Ocean Protocol~\cite{oceanprotocol} has similar goals with Agora, i.e.,~democratizing AI by giving equal opportunities to everyone to access data.
To achieve this they develop a decentralized protocol and network to be used as a foundational substrate to power a new ecosystem of data marketplaces. However, their focus in only on the data aspect.
Datum~\cite{datum} focuses on the privatization and secure storage of data sharing and proposes a network based on blockchain technology that allows users to take control of their data, both personal and data from IoT  devices they own.
Enigma~\cite{enigma} offers a protocol for computations on encrypted data by enabling computational resources to be shared securely in a decentralized manner and
Nebula~\cite{nebula} forms a cloud of edge computers to perform distributed data-intensive computing.
In the space of machine learning, ML Bazaar~\cite{mlBazaar} proposes a unified ML API to ease the development and sharing of ML algorithms. Although such primitives can be used in our specification, Agora goes beyond a simple abstraction to a holistic  solution for democratizing AI and data science.
Although all these efforts are going in the right direction for building a data ecosystem, it is still hard to combine them for devising new solutions.
Our work envisions a single data ecosystem where data, DS technologies, and storage and compute resources can easily be combined to give birth to new data insights or technologies.

There are also initiatives in providing marketplaces for sharing data~\cite{dataspace,dawex}, data science tools~\cite{awsmarketplace,gsuite,azure}, AI~\cite{azure,acumos,awsmarketplace,bonseyes}, and services~\cite{salesforce,experfy}. When it comes to matchmaking, previous solutions are inspired by the semantic web reseach community that address a similar problem for web services~\cite{sws-iswc02}, including solutions for automated web service composition~\cite{web-services}.
The industry has also brought storage, computational, and cloud resources at the reach of the masses.
Amazon EC2~\cite{ec2}, Microsoft Azure~\cite{azure}, and IBM Cloud~\cite{ibmcloud} are just few examples of such efforts.
Nevertheless, all these efforts provide lock-in solutions: Users must stick to one provider for the entire pipeline of their solutions.
We envision an open data ecosystem where one can combine resources from different marketplaces without lock-in effects.

The research community has also proposed many solutions to facilitate data processing in general from different angles:
such as scalable data processing systems~\cite{spark,nephelepact}, declarative data querying~\cite{shark,piglatin}, intelligent systems~\cite{sagedb}, internet-of-things systems~\cite{iot, zeuch2019nebulastream}, and
cross-platform (a.k.a.~polystore) processing~\cite{rheem,bigdawg,musketeer}, among others.
All these works are orthogonal and complementary to our vision: one could see them as the assets being offered in Agora.

  \section{Conclusion}\label{sec:conclusion}
  We presented Agora, our vision towards a unified asset ecosystem.
Assets are fine-grained data-related units of production, such as data, algorithms, and physical infrastructure components.
Agora provides the technical infrastructure for offering, using, and combining assets to form novel data-driven applications and to derive new insights.
One can share assets through marketplaces, use and combine them through asset managers, and execute them through execution managers.
Ultimately, Agora aims at providing open access to the entire data value chain, thereby preventing lock-in effects and removing entry barriers for new asset providers.
We pointed out 15 open research challenges that the database research community should address to make such an asset ecosystem a reality.
We discussed different potential solutions with respect to asset management, complaint asset processing, asset pricing and billing, asset privacy and security, as well as efficient asset execution.

This paper is a call for action as we believe that the database community is well positioned to lead the efforts towards a unified asset ecosystem.
That, in turn, will have positive implications on society, economy, and science:
\begin{packed_item}
  \item {\em Society}:
  It would be used not only by economic operators but also by research institutions, universities, schools, and citizens, which would have a huge benefit in data literacy.
  For example, students could be playfully introduced to programming, data analysis, and even potential business models.
  Lay people could also prepare chores, or even potential business models, by developing on top of the exposed data and analytics infrastructure.
  Most importantly, data and DS technologies could remain with their owners.
  Everyone could contribute to the big asset ecosystem.

  \item {\em Economics}:
  It would provide a breeding ground for data-driven technology innovation by exposing data and DS technologies.
  This would reduce the cost of new insights or the establishment of new business models.
  In this way, it can become an innovation engine for education, business models, business start-ups, and data-driven value creation.
  It would also have a huge impact on small and medium-sized enterprises by having a lower entry threshold for the use of a data and analysis infrastructure.
  For example, it would enable a restaurant to predict how long they will have to stay open on a given evening in order to better plan human resources.
  Additionally, it would motivate a consistent implementation of open standards, which, in turn, could break the current vendor lock-in effects.

  \item {\em Scientific}:
  It would make tools of the entire data value chain (processing, analysis, and visualization) re-usable and easy-to-use (web-based, plug \& play, a combination of public and private data in an analysis).
  This would enable more researchers to derive insights from data without deep knowledge about data management and algorithms.
  It would also foster scientific innovation by enabling researchers to easily share their data insights and technologies.
  Moreover, it would ignite new research in all sciences by providing scientists with access to a large amount of data and state-of-the-art DS technologies.
\end{packed_item}

  \bibliographystyle{abbrv}
  \bibliography{references}

\begin{thebibliography}{100}

\bibitem{seattle-report}
D.~Abadi, A.~Ailamaki, D.~Andersen, P.~Bailis, M.~Balazinska, P.~Bernstein,
  P.~Boncz, S.~Chaudhuri, A.~Cheung, A.~Doan, L.~Dong, M.~J. Franklin,
  J.~Freire, A.~Halevy, J.~M. Hellerstein, S.~Idreos, D.~Kossmann, T.~Kraska,
  S.~Krishnamurthy, V.~Markl, S.~Melnik, T.~Milo, C.~Mohan, T.~Neumann, B.~C.
  Ooi, F.~Ozcan, J.~Patel, A.~Pavlo, R.~Popa, R.~Ramakrishnan, C.~Re,
  M.~Stonebraker, and D.~Suciu.
\newblock {The Seattle Report on Database Research}.
\newblock
  \url{https://db.cs.washington.edu/events/other/2018/Seattle_DBResearch_Report-Full.pdf}.

\bibitem{acumos}
{Acumos}.
\newblock \url{https://www.acumos.org}.

\bibitem{rheem}
D.~Agrawal, S.~Chawla, B.~Contreras-Rojas, A.~Elmagarmid, Y.~Idris, Z.~Kaoudi,
  S.~Kruse, J.~Lucas, E.~Mansour, M.~Ouzzani, P.~Papotti, J.-A. Quiane-Ruiz,
  N.~Tang, S.~Thirumuruganathan, and A.~Troudi.
\newblock {Rheem:} enabling cross-platform data processing -- may the big data
  be with you!
\newblock In {\em PVLDB}, 2018.

\bibitem{ec2}
{Amazon EC2}.
\newblock \url{https://aws.amazon.com/ec2}.

\bibitem{awsmarketplace}
{AWS Marketplace}.
\newblock https://aws.amazon.com/ marketplace.

\bibitem{nephelepact}
D.~Battr{\'e}, S.~Ewen, F.~Hueske, O.~Kao, V.~Markl, and D.~Warneke.
\newblock {Nephele/PACTs}: A programming model and execution framework for
  web-scale analytical processing.
\newblock In {\em SoCC}, 2010.

\bibitem{benson2020Disco}
L.~Benson, P.~M. Grulich, S.~Zeuch, V.~Markl, and T.~Rabl.
\newblock Disco: Efficient distributed window aggregation.
\newblock In {\em EDBT}, 2020.

\bibitem{systemml}
M.~Boehm, M.~W. Dusenberry, D.~Eriksson, A.~V. Evfimievski, F.~M. Manshadi,
  N.~Pansare, B.~Reinwald, F.~R. Reiss, P.~Sen, A.~C. Surve, and S.~Tatikonda.
\newblock {SystemML}: Declarative machine learning on {Spark}.
\newblock In {\em {PVLDB}}, 2016.

\bibitem{bonseyes}
{Bonseyes}.
\newblock \url{https://bonseyes.com/}.

\bibitem{Borkar_2011_The-Future}
S.~Borkar and A.~A. Chien.
\newblock The future of microprocessors.
\newblock {\em Communications of the ACM}, 54(5):67--77, 2011.

\bibitem{Bress_2018_Generating}
S.~Bre{{\ss}}, B.~K{{\"o}}cher, H.~Funke, S.~Zeuch, T.~Rabl, and V.~Markl.
\newblock Generating custom code for efficient query execution on heterogeneous
  processors.
\newblock {\em VLDB Journal}, 27(6):797--822, 2018.

\bibitem{ccpa}
{California consumer privacy act}.
\newblock \url{https://oag.ca.gov/privacy/ccpa}.

\bibitem{carbone2015apache}
P.~Carbone, A.~Katsifodimos, S.~Ewen, V.~Markl, S.~Haridi, and K.~Tzoumas.
\newblock {Apache Flink}: Stream and batch processing in a single engine.
\newblock {\em Bulletin of the IEEE Computer Society Technical Committee on
  Data Engineering}, 36(4), 2015.

\bibitem{carbone2016cutty}
P.~Carbone, J.~Traub, A.~Katsifodimos, S.~Haridi, and V.~Markl.
\newblock Cutty: Aggregate sharing for user-defined windows.
\newblock In {\em CIKM}, 2016.

\bibitem{Chothia2016}
Z.~Chothia, J.~Liagouris, F.~McSherry, and T.~Roscoe.
\newblock Explaining outputs in modern data analytics.
\newblock {\em Proc. VLDB Endow.}, 9(12):1137–1148, Aug. 2016.

\bibitem{corbet2019cryptocurrencies}
S.~Corbet, B.~Lucey, A.~Urquhart, and L.~Yarovaya.
\newblock Cryptocurrencies as a financial asset: A systematic analysis.
\newblock {\em International Review of Financial Analysis}, 62:182--199, 2019.

\bibitem{costan2016intel}
V.~Costan and S.~Devadas.
\newblock Intel {SGX} explained.
\newblock {\em IACR Cryptology ePrint Archive}, (086):1--118, 2016.

\bibitem{DBLP:conf/icde/DashKA09}
D.~Dash, V.~Kantere, and A.~Ailamaki.
\newblock An economic model for self-tuned cloud caching.
\newblock In {\em ICDE}, 2009.

\bibitem{dataspace}
{Data Space}.
\newblock \url{https://www.datapace.io}.

\bibitem{datahub}
{Datahub}.
\newblock \url{https://datahub.io}.

\bibitem{dawex}
{Datawex}.
\newblock \url{https://www.dawex.com}.

\bibitem{dbengines}
{DB Engines}.
\newblock Knowledge base of relational and nosql database management systems.
\newblock \url{https://db-engines.com/}.

\bibitem{denning1996taxonomy}
D.~E. Denning and D.~K. Branstad.
\newblock A taxonomy for key escrow encryption systems.
\newblock {\em Communications of the ACM}, 39(3):34--40, 1996.

\bibitem{bigdawg}
J.~Duggan, A.~J. Elmore, M.~Stonebraker, M.~Balazinska, B.~Howe, J.~Kepner,
  S.~Madden, D.~Maier, T.~Mattson, and S.~Zdonik.
\newblock The {BigDAWG} polystore system.
\newblock {\em SIGMOD Record}, 44(2):11--16, 2015.

\bibitem{web-services}
S.~Dustdar and W.~Schreiner.
\newblock A survey on web services composition.
\newblock {\em International journal of web and grid services}, 1(1):1--30,
  2005.

\bibitem{shark}
C.~Engle, A.~Lupher, R.~Xin, M.~Zaharia, M.~J. Franklin, S.~Shenker, and
  I.~Stoica.
\newblock Shark: Fast data analysis using coarse-grained distributed memory.
\newblock In {\em SIGMOD}, 2012.

\bibitem{enigma}
{Enigma}.
\newblock \url{https://enigma.co}.

\bibitem{experfy}
{Experfy}.
\newblock \url{https://www.experfy.com}.

\bibitem{Fang_2019_In-memory}
J.~Fang, Y.~T.~B. Mulder, J.~Hidders, J.~Lee, and H.~P. Hofstee.
\newblock In-memory database acceleration on {FPGAs}: A survey.
\newblock {\em VLDB Journal}.

\bibitem{theeconomist}
{Fuel of the future}.
\newblock Data is giving rise to a new economy.
\newblock {\em The Economist - The world's most valuable resource - Data and
  the new rules of competition}, 2017.
\newblock \url{https://econ.st/2uRnmOw}.

\bibitem{gsuite}
{G Suite Marketplace}.
\newblock \url{https://gsuite.google.com/marketplace}.

\bibitem{gdpr}
{General data protection regulation}.
\newblock \url{https://gdpr-info.eu/}.

\bibitem{musketeer}
I.~Gog, M.~Schwarzkopf, N.~Crooks, M.~P. Grosvenor, A.~Clement, and S.~Hand.
\newblock Musketeer: all for one, one for all in data processing systems.
\newblock In {\em EuroSys}, 2015.

\bibitem{Gregg_2011_Where}
C.~Gregg and K.~Hazelwood.
\newblock Where is the data? why you cannot debate {CPU} vs. {GPU} performance
  without the answer.
\newblock In {\em ISPASS}, 2011.

\bibitem{datum}
R.~Haenni.
\newblock Datum network - the decentralized data marketplace - white paper v15.
\newblock 2017.
\newblock \url{https://datum.org/assets/Datum-WhitePaper.pdf}.

\bibitem{havet2017securestreams}
A.~Havet, R.~Pires, P.~Felber, M.~Pasin, R.~Rouvoy, and V.~Schiavoni.
\newblock Securestreams: A reactive middleware framework for secure data stream
  processing.
\newblock In {\em DEBS}, pages 124--133, 2017.

\bibitem{He_2009_Relational}
B.~He, M.~Lu, K.~Yang, R.~Fang, N.~K. Govindaraju, Q.~Luo, and P.~V. Sander.
\newblock Relational query coprocessing on graphics processors.
\newblock {\em ACM Transactions on Database Systems}, 34(4):21:1--21:39.

\bibitem{He_2013_Revisiting}
J.~He, M.~Lu, and B.~He.
\newblock Revisiting co-processing for hash joins on the coupled cpu-gpu
  architecture.
\newblock {\em PVLDB}, 6(10):889--900.

\bibitem{he2018sgxlinger}
W.~He, W.~Zhang, S.~Das, and Y.~Liu.
\newblock Sgxlinger: A new side-channel attack vector based on interrupt
  latency against enclave execution.
\newblock In {\em ICCD}, 2018.

\bibitem{4paradigm}
T.~Hey, S.~Tansley, and K.~Tolle.
\newblock {\em The Fourth Paradigm: Data-Intensive Scientific Discovery}.
\newblock 2009.

\bibitem{ibmcloud}
{IBM Cloud}.
\newblock \url{https://www.ibm.com/cloud}.

\bibitem{Idreos_2018_The-Data}
S.~Idreos, K.~Zoumpatianos, B.~Hentschel, M.~S. Kester, and D.~Guo.
\newblock The data calculator: Data structure design and cost synthesis from
  first principles and learned cost models.
\newblock In {\em MOD}, 2018.

\bibitem{iota}
{IOTA Data Marketplace}.
\newblock \url{https://data.iota.org}.

\bibitem{nebula}
A.~Jonathan, M.~Ryden, K.~Oh, A.~Chandra, and J.~Weissman.
\newblock Nebula: Distributed edge cloud for data intensive computing.
\newblock {\em IEEE Transactions on Parallel and Distributed Systems},
  28(11):3229--3242, 2017.

\bibitem{ml4all}
Z.~Kaoudi, J.-A. Quian{\'e}-Ruiz, S.~Thirumuruganathan, S.~Chawla, and
  D.~Agrawal.
\newblock A cost-based optimizer for gradient descent optimization.
\newblock In {\em SIGMOD}, 2017.

\bibitem{Koutris-pods12}
P.~Koutris, P.~Upadhyaya, M.~Balazinska, B.~Howe, and D.~Suciu.
\newblock Query-based data pricing.
\newblock {\em Journal of the ACM (JACM)}, 62(5):1--44, 2015.

\bibitem{sagedb}
T.~Kraska, M.~Alizadeh, A.~Beutel, E.~H. Chi, J.~Ding, A.~Kristo, G.~Leclerc,
  S.~Madden, H.~Mao, and V.~Nathan.
\newblock {SageDB}: A learned database system.
\newblock In {\em CIDR}, 2019.

\bibitem{lee2019keystone}
D.~Lee, D.~Kohlbrenner, S.~Shinde, D.~Song, and K.~Asanovic.
\newblock Keystone: An open framework for architecting tees, 2019.
\newblock arXiv:1907.10119.

\bibitem{Lindholm_2008_NVIDIA}
E.~Lindholm, J.~Nickolls, S.~Oberman, and J.~Montrym.
\newblock {NVIDIA Tesla}: A unified graphics and computing architecture.
\newblock {\em IEEE Micro}, 28(2):39--55, 2008.

\bibitem{lundqvist2017thing}
T.~Lundqvist, A.~de~Blanche, and H.~R.~H. Andersson.
\newblock Thing-to-thing electricity micro payments using blockchain
  technology.
\newblock In {\em 2017 Global Internet of Things Summit (GIoTS)}, pages 1--6.
  IEEE, 2017.

\bibitem{madden2002tag}
S.~Madden, M.~J. Franklin, J.~M. Hellerstein, and W.~Hong.
\newblock Tag: A tiny aggregation service for ad-hoc sensor networks.
\newblock {\em ACM SIGOPS Operating Systems Review}, 36(SI):131--146, 2002.

\bibitem{sourcecode-searchengine}
C.~McMillan, M.~Grechanik, D.~Poshyvanyk, C.~Fu, and Q.~Xie.
\newblock {Exemplar: A Source Code Search Engine for Finding Highly Relevant
  Applications}.
\newblock {\em IEEE Transactions on Software Engineering}, 38(5):1069--1087,
  2012.

\bibitem{recsys19-tutorial}
R.~Mehrotra and B.~Carterette.
\newblock Recommendations in a marketplace (tutorial).
\newblock In {\em ACM Recommender Systems (RecSys)}, 2019.

\bibitem{azure}
{Microsoft Azure}.
\newblock \url{https://azure.microsoft.com}.

\bibitem{iot}
D.~Miorandi, S.~Sicari, F.~De~Pellegrini, and I.~Chlamtac.
\newblock Internet of things: Vision, applications and research challenges.
\newblock {\em Ad Hoc Networks}, 2012.

\bibitem{ray}
P.~Moritz, R.~Nishihara, S.~Wang, A.~Tumanov, R.~Liaw, E.~Liang, M.~Elibol,
  Z.~Yang, W.~Paul, M.~I. Jordan, and I.~Stoica.
\newblock Ray: {A} distributed framework for emerging {AI} applications.
\newblock In {\em OSDI}, pages 561--577, 2018.

\bibitem{morrissey2010tls}
P.~Morrissey, N.~P. Smart, and B.~Warinschi.
\newblock The tls handshake protocol: A modular analysis.
\newblock {\em Journal of Cryptology}, 23(2):187--223, 2010.

\bibitem{Namaki2020}
M.~H. Namaki, A.~Floratou, F.~Psallidas, S.~Krishnan, A.~Agrawal, and Y.~Wu.
\newblock Vamsa: Tracking provenance in data science scripts, 2020.
\newblock arXiv:2001.01861.

\bibitem{navteq}
{Navteq}.
\newblock \url{https://here.navigation.com/}.

\bibitem{ngabonziza2016trustzone}
B.~Ngabonziza, D.~Martin, A.~Bailey, H.~Cho, and S.~Martin.
\newblock Trustzone explained: Architectural features and use cases.
\newblock In {\em IEEE CIC}, 2016.

\bibitem{program-translation}
A.~T. Nguyen, Z.~Tu, and T.~N. Nguyen.
\newblock {Do contexts help in phrase-based, statistical source code
  migration?}
\newblock In {\em ICSME}, 2016.

\bibitem{oceanprotocol}
{Ocean Protocol}.
\newblock \url{https://oceanprotocol.com}.

\bibitem{piglatin}
C.~Olston, B.~Reed, U.~Srivastava, R.~Kumar, and A.~Tomkins.
\newblock {Pig Latin}: A not-so-foreign language for data processing.
\newblock In {\em SIGMOD}, 2008.

\bibitem{openai}
{OpenAI}.
\newblock \url{https://openai.com}.

\bibitem{sws-iswc02}
M.~Paolucci, T.~Kawamura, T.~R. Payne, and K.~Sycara.
\newblock Semantic matching of web services capabilities.
\newblock In {\em International semantic web conference}, pages 333--347.
  Springer, 2002.

\bibitem{park2019streambox}
H.~Park, S.~Zhai, L.~Lu, and F.~X. Lin.
\newblock {StreamBox-TZ}: secure stream analytics at the edge with {TrustZone}.
\newblock In {\em {USENIX} {ATC}}, pages 537--554, 2019.

\bibitem{Patterson2018}
E.~Patterson, I.~Baldini, A.~Mojsilovic, and K.~R. Varshney.
\newblock Teaching machines to understand data science code by semantic
  enrichment of dataflow graphs, 2018.
\newblock arXiv:1807.05691.

\bibitem{comparison}
A.~Pavlo, E.~Paulson, A.~Rasin, D.~J. Abadi, D.~J. DeWitt, S.~Madden, and
  M.~Stonebraker.
\newblock A comparison of approaches to large-scale data analysis.
\newblock In {\em SIGMOD}, 2009.

\bibitem{popov2016tangle}
S.~Popov.
\newblock The tangle.
\newblock {\em cit. on}, 2016.

\bibitem{porter2018introducing}
N.~Porter, J.~Garms, and S.~Simakov.
\newblock Introducing {Asylo}: an open-source framework for confidential
  computing, 2018.
\newblock
  \url{https://cloud.google.com/blog/products/gcp/introducing-asylo-an-open-source-framework-for-confidential-computing}.

\bibitem{priebe2018enclavedb}
C.~Priebe, K.~Vaswani, and M.~Costa.
\newblock {EnclaveDB}: A secure database using {SGX}.
\newblock In {\em IEEE Symposium on Security and Privacy (SP)}, pages 264--278.
  IEEE, 2018.

\bibitem{Publio2018}
G.~C. Publio, D.~Esteves, A.~{\L}awrynowicz, P.~Panov, L.~Soldatova, T.~Soru,
  J.~Vanschoren, and H.~Zafar.
\newblock {ML Schema}: Exposing the semantics of machine learning with schemas
  and ontologies.
\newblock 2018.
\newblock arXiv:1807.05351.

\bibitem{sblq-journal}
J.-A. Quiané-Ruiz, P.~Lamarre, and P.~Valduriez.
\newblock {A Self-Adaptable Query Allocation Framework for Distributed
  Information Systems}.
\newblock {\em VLDB Journal}, 18(3):649--674.

\bibitem{sqlb}
J.-A. Quiané-Ruiz, P.~Lamarre, and P.~Valduriez.
\newblock {SQLB: A Query Allocation Framework for Autonomous Consumers and
  Providers}.
\newblock In {\em VLDB}, 2007.

\bibitem{sbqa}
J.-A. Quiané-Ruiz, P.~Lamarre, and P.~Valduriez.
\newblock {SbQA: A Self-Adaptable Query Allocation Process.}
\newblock In {\em ICDE}, 2009.

\bibitem{Redyuk2019a}
S.~Redyuk.
\newblock Automated documentation of end-to-end experiments in data science.
\newblock In {\em ICDE}, 2019.

\bibitem{Rosenfeld_2019_Performance}
V.~Rosenfeld, S.~Bre{{\ss}}, S.~Zeuch, T.~Rabl, and V.~Markl.
\newblock Performance analysis and automatic tuning of hash aggregation on
  gpus.
\newblock In {\em DaMoN}, 2019.

\bibitem{Rosenfeld_2015_The-Operator}
V.~Rosenfeld, M.~Heimel, C.~Viebig, and V.~Markl.
\newblock The operator variant selection problem on heterogeneous hardware.
\newblock In {\em ADMS}, 2015.

\bibitem{salesforce}
{Salesforce}.
\newblock \url{https://www.salesforce.com}.

\bibitem{Schelter2018}
S.~Schelter, F.~Biessmann, T.~Januschowski, D.~Salinas, S.~Seufert, G.~Szarvas,
  M.~Vartak, S.~Madden, H.~Miao, A.~Deshpande, et~al.
\newblock On challenges in machine learning model management.
\newblock {\em IEEE Data Engineering}, 2018.

\bibitem{Schelter2017}
S.~Schelter, J.-H. B{\"o}se, J.~Kirschnick, T.~Klein, and S.~Seufert.
\newblock Automatically tracking metadata and provenance of machine learning
  experiments.
\newblock {\em Machine Learning Systems workshop at NIPS}, 2017.

\bibitem{klausschwab}
K.~Schwab.
\newblock Mastering the fourth industrial revolution.
\newblock {\em Foreign Affairs}, 2015.

\bibitem{mlBazaar}
M.~J. Smith, C.~Sala, J.~M. Kanter, and K.~Veeramachaneni.
\newblock The machine learning bazaar: Harnessing the ml ecosystem for
  effective system development.
\newblock 2019.
\newblock arXiv:1905.08942.

\bibitem{mariposa}
M.~Stonebraker, P.~M. Aoki, W.~Litwin, A.~Pfeffer, A.~Sah, J.~Sidell,
  C.~Staelin, and A.~Yu.
\newblock {Mariposa: A Wide-Area Distributed Database System}.
\newblock {\em {VLDB} Journal}, 5(1):48--63, 1996.

\bibitem{sunyaev2013cloud}
A.~Sunyaev and S.~Schneider.
\newblock Cloud services certification.
\newblock {\em Communications of the ACM}, 56(2):33--36, 2013.

\bibitem{tang2017clkscrew}
A.~Tang, S.~Sethumadhavan, and S.~Stolfo.
\newblock {CLKSCREW}: exposing the perils of security-oblivious energy
  management.
\newblock In {\em {USENIX} Security Symposium}, 2017.

\bibitem{Teubner_2013_Data}
J.~Teubner and L.~Woods.
\newblock Data processing on {FPGAs}.
\newblock {\em Synthesis Lectures on Data Management}, 5(2):1--118, 2013.

\bibitem{oxfordeconomics}
{The future of data: Adjusting to an opt-in economy}.
\newblock {\em Oxford Economics Report}, 2018.
\newblock
  \url{https://www.oxfordeconomics.com/recent-releases/the-future-of-data}.

\bibitem{thoma2019behind}
C.~Thoma, A.~J. Lee, and A.~Labrinidis.
\newblock Behind enemy lines: Exploring trusted data stream processing on
  untrusted systems.
\newblock In {\em ACM Conference on Data and Application Security and Privacy},
  pages 243--254, 2019.

\bibitem{traub2018scotty}
J.~Traub, P.~M. Grulich, A.~R. Cuellar, S.~Bre{\ss}, A.~Katsifodimos, T.~Rabl,
  and V.~Markl.
\newblock Scotty: Efficient window aggregation for out-of-order stream
  processing.
\newblock In {\em ICDE}, 2018.

\bibitem{traub2019efficient}
J.~Traub, P.~M. Grulich, A.~R. Cu{\'e}llar, S.~Bre{\ss}, A.~Katsifodimos,
  T.~Rabl, and V.~Markl.
\newblock Efficient window aggregation with general stream slicing.
\newblock In {\em EDBT}, 2019.

\bibitem{thecase}
D.~Tsoumakos and C.~Mantas.
\newblock The case for multi-engine data analytics.
\newblock In {\em Euro-Par}, 2013.

\bibitem{datalawyer}
P.~Upadhyaya, M.~Balazinska, and D.~Suciu.
\newblock Automatic enforcement of data use policies with datalawyer.
\newblock In {\em Proceedings of the 2015 ACM SIGMOD International Conference
  on Management of Data}, SIGMOD ’15, page 213–225, 2015.

\bibitem{vinayagamurthy2019stealthdb}
D.~Vinayagamurthy, A.~Gribov, and S.~Gorbunov.
\newblock {StealthDB}: a scalable encrypted database with full {SQL} query
  support.
\newblock {\em Proceedings on Privacy Enhancing Technologies},
  2019(3):370--388, 2019.

\bibitem{wang2017enabling}
J.~Wang, Z.~Hong, Y.~Zhang, and Y.~Jin.
\newblock Enabling security-enhanced attestation with intel sgx for remote
  terminal and iot.
\newblock {\em IEEE Transactions on Computer-Aided Design of Integrated
  Circuits and Systems}, 37(1):88--96, 2017.

\bibitem{xu2016blockchain}
A.~Xu, M.~Li, X.~Huang, N.~Xue, J.~Zhang, and Q.~Sheng.
\newblock A blockchain based micro payment system for smart devices.
\newblock {\em Signature}, 256(4936):115, 2016.

\bibitem{yelp}
{{Y}elp display requirements}.
\newblock \url{https://www.yelp.com/dataset/}.

\bibitem{Yuan_2013_The-Yin-and-Yang}
Y.~Yuan, R.~Lee, and X.~Zhang.
\newblock The yin and yang of processing data warehousing queries on {GPU}
  devices.
\newblock In {\em PVLDB}, 2013.

\bibitem{spark}
M.~Zaharia, M.~Chowdhury, M.~J. Franklin, S.~Shenker, and I.~Stoica.
\newblock Spark: Cluster computing with working sets.
\newblock In {\em {USENIX HotCloud}}, 2010.

\bibitem{zaharia2016apache}
M.~Zaharia, R.~S. Xin, P.~Wendell, T.~Das, M.~Armbrust, A.~Dave, X.~Meng,
  J.~Rosen, S.~Venkataraman, M.~J. Franklin, et~al.
\newblock {Apache Spark}: a unified engine for big data processing.
\newblock {\em Communications of the ACM}, 59(11):56--65, 2016.

\bibitem{zeuch2019nebulastream}
S.~Zeuch, A.~Chaudhary, B.~Del~Monte, H.~Gavriilidis, D.~Giouroukis, P.~M.
  Grulich, S.~Bre{\ss}, J.~Traub, and V.~Markl.
\newblock The {NebulaStream Platform}: Data and application management for the
  internet of things.
\newblock {\em CIDR}, 2020.

\bibitem{zheng2017opaque}
W.~Zheng, A.~Dave, J.~G. Beekman, R.~A. Popa, J.~E. Gonzalez, and I.~Stoica.
\newblock Opaque: An oblivious and encrypted distributed analytics platform.
\newblock In {\em {USENIX} {NSDI}}, pages 283--298, 2017.

\end{thebibliography}
  \normalsize
\end{document}